\documentclass[12pt,a4paper]{article}
\usepackage{amsmath,amssymb,citesort,enumerate}
\usepackage{epsfig}
\usepackage{hhline,multirow}

\newcommand{\eqeqref}[1]{Eq.~\eqref{#1}}
\newcommand{\eqseqref}[1]{Eqs.~\eqref{#1}}
\newcommand{\refref}[1]{Ref.~\cite{#1}}
\newcommand{\refsref}[1]{Refs.~\cite{#1}}
\newcommand{\secref}[1]{Sec.~\ref{#1}}
\newcommand{\appref}[1]{App.~\ref{#1}}
\newcommand{\figref}[1]{Fig.~\ref{#1}}
\newcommand{\tabref}[1]{Table~\ref{#1}}

\newcommand{\beq}{\begin{equation}}
\newcommand{\eeq}{\end{equation}}
\newcommand{\bea}{\begin{eqnarray}}
\newcommand{\beas}{\begin{eqnarray*}}
\newcommand{\beau}[1]{\begin{equation} \label{#1} \begin{array}{rcl}}
\newcommand{\eea}{\end{eqnarray}}
\newcommand{\eeas}{\end{eqnarray*}}
\newcommand{\eeau}{\end{array} \end{equation}}
\newcommand{\bay}{\begin{array}}
\newcommand{\eay}{\end{array}}
\newcommand{\bals}{\begin{align*}}
\newcommand{\eals}{\end{align*}}

\newcommand{\ds}{\displaystyle}

\newcommand{\lora}{{\longrightarrow}}

\newcommand{\ra}{{\rightarrow}}

\newcommand{\vev}[1]{\langle #1 \rangle}

\newcommand{\esp}[1]{\, e^{\,\,\textstyle {#1}}}
\newcommand{\In}[2]{ \left. #1 \right| _{#2} }

\begin{document}



\begin{titlepage}
\title{\LARGE\bf Hadron production \\
in deep inelastic lepton-nucleus scattering}

\author{
  {\bf A.~Accardi}$^{\,1}$\!
        \thanks{E-mail address: accardi@tphys.uni-heidelberg.de}\;,  
  {\bf V.~Muccifora}$^{\,2}$\!
        \thanks{E-mail address: muccifora@lnf.infn.it}\;,
  {\bf H.J.~Pirner}$^{\,1\,3}$\!
        \thanks{E-mail address: pir@tphys.uni-heidelberg.de}
}

\date{}

\maketitle

\begin{center}
\vskip-1cm
\parbox{13cm}{
\begin{itemize}
\item[$^1$] 
Institut f\"ur Theoretische Physik der Universit\"at Heidelberg \\
Philosophenweg 19, D-69120 Heidelberg, Germany 
\item[$^2$] 
INFN, Laboratori Nazionali di Frascati, I-00044 Frascati, Italy 
\item[$^3$] 
Max-Planck-Institut f\"ur Kernphysik, Heidelberg, Germany
\end{itemize}
}
\end{center}

\vspace{1cm}
\begin{abstract}
Predictions for semi-inclusive deep inelastic lepton-nucleus
scattering are presented. Both the effects of gluon radiation 
by the struck quark and the absorption of the produced hadron are
considered.
The gluon radiation covers a  larger window in virtuality $Q^2$ because of 
the  increased deconfinement of  quarks inside nuclei. 
The absorption of hadrons formed inside the nucleus is 
described with a flavor dependent cross section. 
Calculations for  rescaled fragmentation functions and nuclear absorption 
are compared with the  EMC and HERMES data for N, Cu and Kr targets
with respect to the  deuteron target. 
Predictions for  Ne and Xe targets  in the HERMES kinematic
regime are given.
\end{abstract}

\begin{footnotesize}
\begin{center}
\end{center}
\end{footnotesize}


\end{titlepage}

\newpage 

\setcounter{page}{2}
\setcounter{footnote}{0}

\section{Introduction}
\setcounter{equation}{0}

The fragmentation of  quarks  into hadrons is still not completely understood.
In fact, the final stages of hadronization involve low scale nonperturbative
processes in QCD in Minkowski space together with hadronic wave functions
which are not yet calculable. The  nucleus may help 
to understand the space time evolution of a parton as  the nucleons inside the
nucleus play the role of very nearby detectors of the propagating object. 

\begin{figure}[t]
\begin{center}
\begin{minipage}[t]{16cm}
\begin{center}
\parbox{7.5cm}{\vskip-.5cm
\epsfig{figure=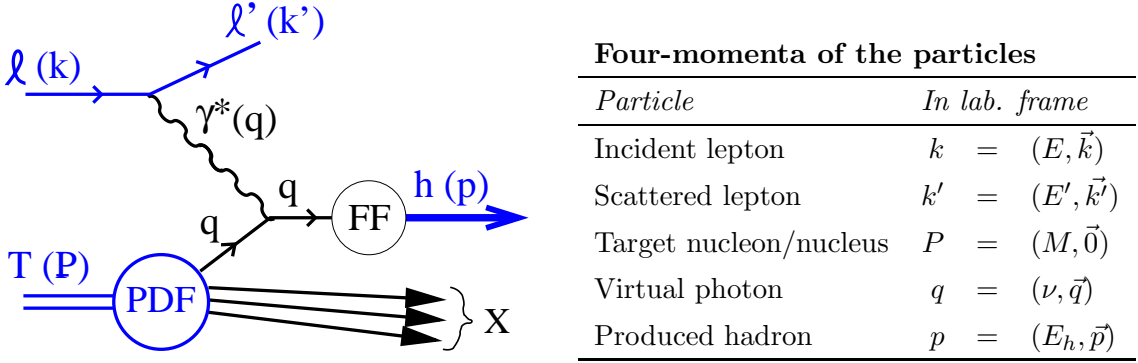,width=7cm}}
\parbox{8cm}{
\small
\begin{tabular}{lrcl} 
\multicolumn{4}{l}{\bf Four-momenta of the particles} \\
\hline 
  \it Particle 
            & \multicolumn{3}{l}{\it In lab. frame}   \\\hline
  Incident lepton 
            & $k$ & = & $(E,\vec{k})$ \\
  Scattered lepton 
            & $k'$ & = & $(E',\vec{k'})$ \\
  Target nucleon/nucleus
            & $P$     & = & $(M,\vec{0})$ \\
  Virtual photon 
            & $q$     & = & $(\nu,\vec{q})$ \\
  Produced hadron  
            & $p$     & = & $(E_h,\vec{p})$ \\\hline
\end{tabular} }
\vskip0cm
\caption{\footnotesize 
Semi-inclusive hadron production in deep inelastic
scattering on a target T in the pQCD factorization approach. Parton 
distribution functions (PDF) and fragmentation
functions (FF) represent the non-perturbative
input. Four-momenta of the particles involved in the
process in the laboratory frame are defined in the table on the right.}
\label{fig:DIS}
\end{center}
\end{minipage}
\end{center}
\vskip-.4cm
\end{figure}

Deep inelastic lepton-nucleus scattering has the advantage that the
lepton transfers a well-defined energy $\nu$ to the struck quark 
propagating through cold nuclear matter. 
Due to factorization in deep inelastic scattering (DIS), the semi-inclusive
cross section can be described by the product of a parton distribution function
(PDF)
with  a fragmentation function (FF) and the hard scattering cross section.
\figref{fig:DIS} shows a schematic diagram of semi-inclusive
deep inelastic lepton scattering on a target, and the definitions
of the four momenta of the particles involved in the process.
In the naive parton model, the probability  $q_f(x)$ 
that a  quark of flavor $f$
with momentum fraction $x$  
is present in the target is multiplied with  the probability  $D_f^h(z)$
that it hadronizes 
into a definite hadron $h$ which carries a momentum fraction $z$ of 
the quark.

While the inclusive deep inelastic scattering of high 
energy leptons on nuclei has been used to measure the 
quark distributions $q_f(x)$ in nuclei,
semi-inclusive DIS (SIDIS) can be used to study 
the medium modifications of quark fragmentation  functions $D_f^h(z)$.
In SIDIS, besides the scattered lepton $l\,'$,
the leading hadron $h$ formed from the struck quark is detected 
with energy $E_h=z\nu$ in the target rest frame (see
\tabref{tab:DISkinvar} for a list of kinematic variables).
Experimentally, hadron multiplicities 
are obtained from
normalizing the SIDIS yield $N^h$ to the DIS yield $N^{DIS}$: 
$({1}/{N ^{DIS}}) ({dN^h(x,z)}/{dz}) \approx 
  ({ \sum_f e_f^2 {q_f(x)} {D_f^h(z)}}/
         {\sum_f e_f^2 {q_f(x)}})$, where the sum over all the quark of a flavor $f$ and charge $e_f$ is performed.
It has been shown that, after integrating over a broad range of
$x$, the multiplicities are a good approximation to the 
fragmentation functions (see \refref{ai1}
and references therein). Thus, a measurement of the hadron multiplicity 
in nuclei
is a good tool to measure possible medium modifications of quark 
fragmentation functions.

The experimental results on semi-inclusive leptoproduction of hadrons 
from nuclei \cite{EMC,HERMES}
are usually presented in terms of multiplicity ratios $R_M^h$ 
between  nuclear  ($A$) and deuteron ($D$) targets, as
functions of $z$ and $\nu$ respectively:
\begin{align} 
    R_M^h(z) & = 
        \frac{\ds 1}{\ds N_A^{DIS}} \frac{\ds dN_A^h}{\ds dz}
        \bigg/ \frac{\ds 1}{\ds N_D^{DIS}} \frac{\ds dN_D^h}{\ds dz} 
        \, 
 \label{multratioz} \\
    R_M^h(\nu) & = 
        \frac{\ds 1}{\ds N_A^{DIS}} \frac{\ds dN_A^h}{\ds d\nu}
        \bigg/ \frac{\ds 1}{\ds N_D^{DIS}} \frac{\ds dN_D^h}{\ds d\nu} 
        \, 
 \label{multrationu}
\end{align}
 
In absence of nuclear effects, the ratio $R^h_M$ would be equal to
1, and experimental results show that this is the case at high transfer energy 
$\nu$ \cite{EMC}. 

The HERMES experiment \cite{HERMES} has brought new insight into the study of the 
hadronization process, since
it has access to
lower energy transfers $\nu$, where the hadronization is supposed to
occur inside the nucleus. It has also extended the
range of fractional momenta  to  larger values of $z$. 
In principle, a comparison with the EMC results \cite{EMC}, which  
involve  much higher  transfer energy, would allow us 
to separate the physics of gluon radiation from the absorption
process of the produced hadron.

There has been an intensive discussion on energy loss of partons in
a cold and hot nuclear environment \cite{enloss}, which then leads to modified
fragmentation functions in nuclei or in deconfined quark matter
\cite{modfrag}.
Recent data on high transverse momentum hadron produced in Au-Au
collisions at RHIC have truly opened up the domain of hard scattering
and jet-quenching in heavy-ion collisions \cite{RHICdata}.
We would like to address this problem from a different perspective.
Already in the vacuum, the struck quark loses substantial energy in
the gluon cascade; the question arises how this cascade process is
modified in the nuclear medium.

The goal of this paper is to describe the $R^h_M$ measurements 
\cite{EMC,HERMES}\footnote{The SLAC data \cite{SLAC} will not be
considered in this paper since the semi-inclusive distributions have
been normalized to the target density rather than to the rate of
inclusive events. Therefore they  can not be  
directly compared to EMC and HERMES data.}
in terms of modified fragmentation functions 
supplemented by nuclear absorption.
We describe the modification of
the structure function and fragmentation function in the framework
of rescaling models \cite{NP84,CJRR84,DDD86}.  These models have been
successfully applied to the interpretation of nuclear structure functions. 
In this paper, we extend
the  rescaling model also to the medium
modification of the quark fragmentation functions. 
For the nuclear absorption, we work out two specific fragmentation models 
based on work by
Bialas-Gyulassy \cite{BG87} and Bialas-Chmaj \cite{BC83}.
Applications of the above mentioned nuclear absorption models to DIS
on nuclear targets may be found in \refsref{CS92} and \cite{C89}.

Recently, other ``QCD inspired'' analysis such as
the gluon-bremsstrahlung calculation 
for leading hadron production \cite{KNP96},
the higher-twist pQCD computation of  \refref{hightwist}
and the effect of parton energy loss  of \refref{ARLEO}
have been applied to hadron production in DIS on nuclei.
Computations for meson production in DIS on 
nuclei in terms of stationary string model are also presented
in \refref{AEG02}.

The outline of the paper  is as follows:
in \secref{sec:rescaling}, we describe the modification of the fragmentation 
function by gluon radiation
of the struck quark; in \secref{sec:probdist},
we discuss the formation length and  
in \secref{sec:nuclearabsorption}, the nuclear absorption of the produced
hadron.
The  comparison between the model predictions and 
the experimental data for charged hadrons is made in
\secref{sec:chargedhadrons}.
The flavor dependence of the hadronization process is 
calculated and compared with data in \secref{sec:flavours}. 
Conclusions are reported in \secref{sec:conclusions}.
\begin{table}[tb]
\small
\begin{center}
\begin{tabular}{rclcll} 
\multicolumn{4}{l}{\bf Kinematic variables} \\
\hline 
  \it Variable & & \multicolumn{2}{l}{\it Covariant} & {\it
    Lab. frame} & \\\hline
  $Q^2$ & = & $-q^2$ & $\simeq$ & $2Mx\nu$ 
        & \parbox[t]{6cm}{Negative four-momentum squared of the
    virtual photon.}\\
  $\nu$ & = & $\frac{q\cdot p}{\sqrt{P^2}}$ & = & $E'-E$ 
        & \parbox[t]{6cm}{Energy of the virtual photon in lab. frame.}\\
  $x$ & = & $\frac{-q^2}{2P\cdot q}$ & = & $\frac{Q^2}{2M\nu}$ 
        & \parbox[t]{6cm}{Bjorken scaling variable (fraction of $P$
          carried by the struck parton in the Breit frame).}\\
  $z$   & = & $\frac{p\cdot P}{q\cdot P}$ & = & $\frac{E_h}{\nu}$ 
        & \parbox[t]{6cm}{Fraction of the virtual photon energy
          carried by the  hadron.}\\
  $y$ & = & $\frac{q\cdot P}{k\cdot P}$ & = & $\frac{\nu}{E}$ 
        & \parbox[t]{6cm}{Fraction of the incident lepton energy
          transferred to the target.}\\
  $W^2$ & = & $(P+q)^2$ & = & $M^2 + 2M\nu - Q^2$ 
        & \parbox[t]{6cm}{Invariant mass squared of the total hadronic
          final state.
          \vskip.2cm}\\\hline
\end{tabular} \\
\parbox[t]{15cm}{\vskip0cm
\caption{\footnotesize
Definitions of the kinematic variables
used in the paper.} 
\label{tab:DISkinvar}}
\end{center}
\end{table}

\section{Modification of parton distribution and fragmentation
  functions by gluon radiation.}
\setcounter{equation}{0}
\label{sec:rescaling}

It has been shown (for a recent review see \refref{Strikman}),
that a simple folding of nucleon structure functions 
with the momentum distribution of nucleons in the nucleus is
insufficient to reproduce the nuclear structure functions.
Nuclear structure and fragmentation functions certainly depend on the
density of baryonic matter and on the amount of momentum sharing
induced by the close packing of nucleons. 
A good description of nuclear structure functions has been achieved 
in the so-called deconfinement models, 
based on the hypothesis that quarks in bound nucleons have access to a
larger region in space than in free nucleons \cite{NP84,CJRR84,NP87}.

Both parton distribution functions and fragmentation functions
depend on the virtuality $Q^2$ of the DIS process.
Their adjustment to the physical scale $Q^2$ takes into account all 
radiated gluons before and after the photon-quark interaction in 
the leading logarithm approximation.  
The long wavelength spectrum  of gluons extends farther into
the infrared toward low $Q^2 \propto 1/\lambda^2$ where $\lambda$ is
the confinement scale. 
Deconfinement models assume a larger
confinement scale $\lambda_A$ in nuclei, compared with the confinement
scale $\lambda_0$ in free nucleons: 
\begin{align*}
        \lambda_A > \lambda_0 \ .
\end{align*}
Therefore, in nuclei gluon radiation would be affected
by the assumed deconfinement of color.

Take, e.g., a parton distribution function and consider
a quark which carries a momentum $Q_0$ when it is 
confined on a scale $\lambda_0$. If the scale changes to  $\lambda_A$ 
it carries a corresponding momentum $Q_A$.
If the free and bound nucleon would  be characterized by these scales
alone, the structure function of the bound nucleon would be 
related to the structure function of the free nucleon by
replacing the measured $Q^2$ by  
$(\lambda_A/\lambda_0)^2 Q^2$ in the arguments of the structure
and fragmentation functions.
Taking into account also the running of the QCD coupling, 
i.e. the existence of the QCD scale $\Lambda_{QCD}$, 
the correct rescaling factor is \cite{NP87,CJRR84}:
\begin{align}
  \xi_A(Q^2) = \Big(\frac{\lambda_A}{\lambda_0}\Big)
                ^{2 \frac{\bar{\alpha_s}}{\alpha_s(Q^2)}} \ ,
\end{align}
where $\alpha_s(Q^2)$ is computed at leading order with 
$\Lambda_{\rm QCD}=200$ MeV and four quark flavours. 
The real scale factor $\xi_A(Q^2)$ 
considers that the low lying modes couple with 
a coupling constant $\bar \alpha_s$.
Effectively, the DGLAP evolution of 
the nuclear structure function 
covers a larger interval in momentum compared with 
the corresponding functions in the nucleon at the same scale Q, 
and the nuclear modified distribution function 
$q_f^A$ for a quark of flavour $f$ reads:
\begin{align}
   q^A_f(x,Q^2) & = q_f(x,\xi_A(Q^2) Q^2) \ ,
 \label{rescPDF} 
\end{align}
where $q_f$ is the corresponding distribution function in a free nucleon.
For consistency it is necessary that partial deconfinement in nuclei 
does not only modify  the parton distribution functions  but also
the fragmentation functions \cite{DDD86}. 
A procedure similar to the above one for the fragmentation function
gives  
\begin{align}
   D^{h|A}_f(z,Q^2) & = D^h_f(z,\xi_A(Q^2) Q^2) \ ,
 \label{rescFF}
\end{align}
where $D_f^h$ is the fragmentation function of a quark of flavour $f$
into a hadron $h$ and $D_f^{h|A}$ is the nuclear modified
fragmentation function.
Of course such a simple adjustment of evolution scale is only a simple 
prescription to mimic the effects of the nucleus on the light-cone 
wave function of the bound nucleon.
The rescaling relations \eqref{rescPDF} and \eqref{rescFF} 
are expected approximately to be valid for $0.1<x<0.6$ and $z\geq0.1$.

We use two models for the computation of the scale factor.
The maximal deconfinement model (MD) \cite{NP84,NP87}
assumes the onset of ``colour conductivity'' in nuclei and takes
$\lambda_A=R_A$, where $R_A$ is the nuclear radius, so that  
\begin{align}
        \frac{\lambda_A}{\lambda_0} = \frac{R_A}{R_p} \ ,
 \label{rescNP} 
\end{align}
Since in this model the low lying modes can become very soft 
$\bar\alpha_s$ is frozen at $\bar\alpha_s=0.54$.
The partial deconfinement model (PD) \cite{CJRR84} 
assumes the deconfinement scale $\lambda_A$ to be proportional to the
overlap of nucleons inside the given nucleus, and $\bar\alpha_s =
\alpha_s(\mu_A^2)$, where $\mu_A = \frac{\lambda_0}{\lambda_A} \mu_0$ and 
$\mu_0=0.66$ GeV$^2$ is the starting scale for DGLAP 
evolution in the free nucleon. 
A recent analysis \cite{SSS02}, based on the 
experimental results \cite{reduceddeconf} 
on the nucleon form factor in nuclei, found that 
the amount of deconfinement predicted by \refref{CJRR84} is too large.
In particular  \refref{SSS02} gives 
an upper limit for $(\lambda_A-\lambda_0)/\lambda_0$$\sim$10\% on iron,
while \refref{CJRR84} predicts a 15.3\% enhancement. 
Therefore following \refref{SSS02}
we reduce the $\lambda_A$ computed in \refref{CJRR84} 
by a universal factor:
\begin{align}
    \left( \frac{\lambda_A}{\lambda_0} - 1 \right) = 
        \frac{0.10}{0.153} 
        \left( \In{\frac{\lambda_A}{\lambda_0}}{CJRR} - 1 \right) \ .
 \label{redresc}
\end{align}
In \tabref{tab:scfact} we show the values of the scale factor $\xi_A$ 
predicted by  the aforementioned maximal deconfinement (MD) and partial deconfinement (PD) models for several nuclei.

\begin{table}[tb]
\small
\begin{center}
\begin{minipage}[t]{16cm}
\begin{center}
\begin{tabular}{|l|c|ccccccc|} 
\hline 
  & & $^2$D$_1$ & $^{14}$N$_7$ & $^{20}$Ne$_{10}$ & $^{56}$Fe$_{26}$ 
    & $^{63}$Cu$_{29}$ & $^{84}$Kr$_{36}$ & $^{131}$Xe$_{54}$
    \\ \hline
  \multirow{2}{1.8cm}{PD model}
    & $\lambda_A/\lambda_0$
    & 1.010 & 1.071 & 1.068 & 1.100 & 1.101 & 1.110 & 1.118 \\ 
  &
    $\sqrt{\xi_{A}}${\scriptsize ($Q=1.5$ GeV)} 
    & 1.014 & 1.110 & 1.106 & 1.164 & 1.165 & 1.182 & 1.195 \\ \hline
  \multirow{2}{1.8cm}{MD model}
    & $\lambda_A/\lambda_0$  
    & 2.46 & 2.99 & 3.53 & 4.42 & 4.62 & 4.88 & 5.68 \\
  &
    $\sqrt{\xi_{A}}${\scriptsize ($Q=1.5$ GeV)} 
    & 3.65 & 4.82 & 6.13 & 8.47 & 9.02 & 9.76 & 12.3\ \ \ \   \\ \hline
\end{tabular}
\caption{\footnotesize
Values of the deconfinement ratio $\lambda_A/\lambda_0$ and scale factors 
$\xi_{A}$ used in partial deconfinement (PD) model (see \eqeqref{redresc}) and
maximal  deconfinement (MD) model (see \eqeqref{rescNP}.} 
\label{tab:scfact}
\end{center}
\end{minipage}
\end{center}
\vskip-.4cm
\end{table}

\begin{figure}[thb]
\begin{center}
\begin{minipage}[t]{15.5cm}
\begin{center}
\parbox[c]{7.5cm}{
\epsfig{figure=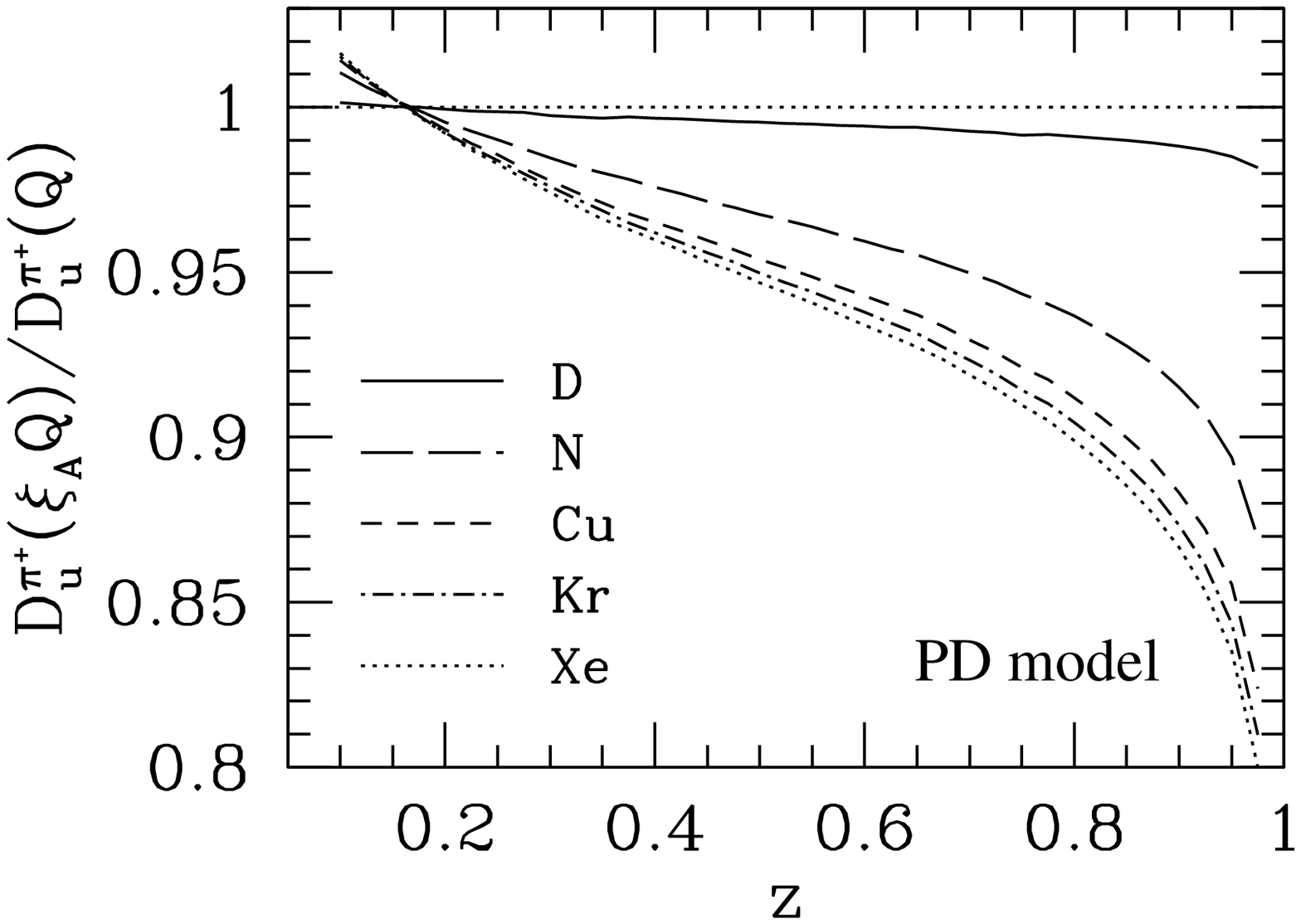,width=7.5cm
,clip=,bbllx=0pt,bblly=150pt,bburx=500pt,bbury=570pt}}
\parbox[c]{7.5cm}{
\epsfig{figure=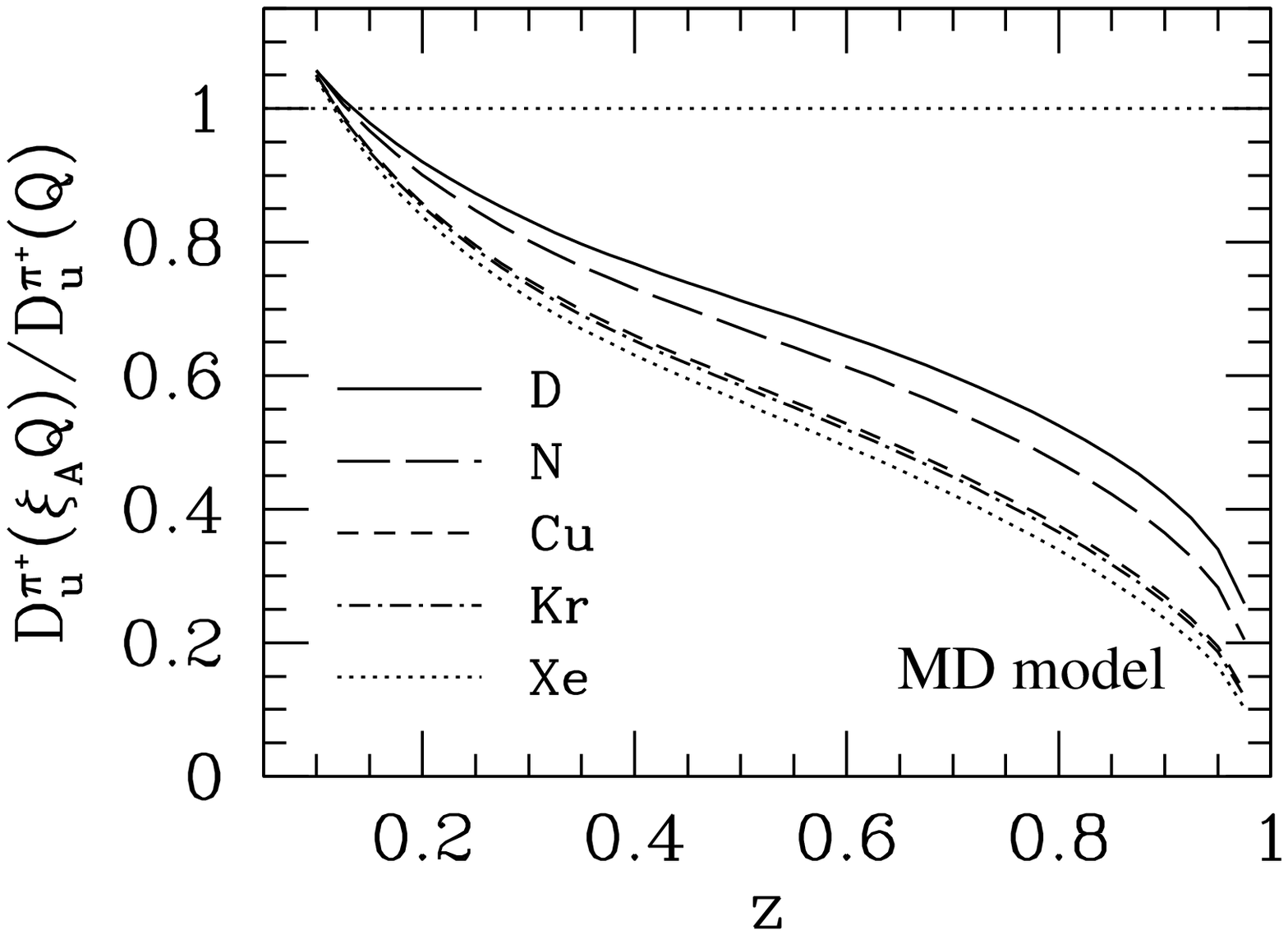,width=7.5cm
,clip=,bbllx=0pt,bblly=150pt,bburx=500pt,bbury=570pt}
}
\vskip-.2cm
\caption{\footnotesize
Ratio of the rescaled fragmentation functions  $u \rightarrow \pi^+$ 
to the standard ones at leading order \cite{K} for $Q^2=2.25$ GeV$^2$ 
as a function of $z$, in the case of partial deconfinement rescaling
(PD) and maximal deconfinement rescaling (MD).}
\label{fig:resconFF}
\end{center}
\end{minipage}
\end{center}
\vskip-.4cm
\end{figure}

The effect of the rescaling on the fragmentation function 
$D_u^{\pi^+}(z,\xi_A Q^2)$  \cite{K} for different nuclei is shown 
in \figref{fig:resconFF} 
both for the partial and maximal deconfinement models.
Note that the MD model  assumes a much larger scale factor, which
results in a larger 
nuclear modification of hadron production especially at high $z$.
One also sees that in the partial deconfinement model the effect of
rescaling on the deuteron is much smaller than on heavy nuclei. 
This is different from the maximal deconfinement model, where only
the nuclear radius matters and rescaling effects gradually vary  
between light and heavy nuclei.
It is also worth to point out that the fragmentation function 
ratios shown in \figref{fig:resconFF} 
increase in the low-$z$ region up to values larger
than unity approximately at $z\sim 0.2$ for all  nuclei in both
rescaling models. This increase, common to all fragmentation
functions with rescaled $Q^2$, 
comes from enhanced gluon radiation at 
small $z$ contributing to hadron production.

We calculate the multiplicity ratio of \eqseqref{multratioz} and 
\eqref{multrationu} by using the rescaled PDF \eqref{rescPDF}
and the rescaled FF \eqref{rescFF} in the leading order pQCD 
computation of $N_A^{DIS}$, $dN_A^h/dz$  and $dN_A^h/d\nu$:
\begin{align}
    \frac{1}{N_A^{DIS}}\frac{dN_A^h}{dz} & = \frac{1}{\sigma^{\ell A}} 
         \int_{\rm exp.\ cuts}\hspace*{-1.0cm}
         dx\, d\nu \sum_f e^2_f\, q_f(x,\xi_A\,Q^2) \,
         \frac{d\sigma^{\ell q}}{dx\,d\nu} \, D_f^{h}(z,\xi_A\,Q^2)  
 \label{dnhdz} \\
    \frac{1}{N_A^{DIS}}\frac{dN_A^h}{d\nu} & = \frac{1}{\sigma^{\ell A}} 
         \int_{\rm exp.\ cuts}\hspace*{-1.0cm}
         dx\, dz \sum_f e^2_f\, q_f(x,\xi_A\,Q^2) \,
         \frac{d\sigma^{\ell q}}{dx\,d\nu} \, D_f^{h}(z,\xi_A\,Q^2)  
 \label{dnhdnu} \\
    \sigma^{\ell A} & = \int_{\rm exp.\ cuts}\hspace*{-1.0cm}
         dx\, d\nu \sum_f e^2_f\, q_f(x,\xi_A\,Q^2) \,
         \frac{d\sigma^{\ell q}}{dx\,d\nu} \ . 
 \label{DISxsec}
\end{align}
Here 
$d\sigma^{\ell q}/dx\,d\nu$ is the differential cross-section for
lepton-quark scattering computed at leading order (LO) in perturbation theory
\cite{PS95}:
\begin{align}
  \frac{d\sigma^{\ell q}}{dx\,d\nu} & = 
    M x \frac{4\pi \alpha^2(Q^2)}{Q^4}\big[1+(1-y)^2\big].
  \nonumber
\end{align}
 Note that the experimental
acceptance is explicitly accounted for 
in the integration limits in \eqseqref{dnhdz}-\eqref{DISxsec}. The
experimental acceptances of the EMC and
HERMES experiments are given in 
\appref{app:EMCandHERMES}. 
Nuclear isospin asymmetry is taken into account by using the 
averaged quark distribution function
\begin{align*}
    q_f^{A} = \frac{1}{A} \big[ Z \, q_f^p + (A-Z) \, q_f^n \big] \ .
\end{align*}
The quark distribution function in the proton and in the
neutron, respectively, $q_f^p$ and $q_f^n$, are related by isospin
symmetry.
In the numerical computations
we use the leading order GRV98 parton distribution 
functions \cite{GRV98}.  
Since the PDF's appear both in the
numerator and in the denominator of \eqseqref{dnhdz}-\eqref{dnhdnu},
the theoretical uncertainty resulting from a different 
parametrization of the distribution functions is minimal. 
For the same reason, the effect of 
shadowing corrections to PDF's in the small-x region is negligible
(the lowest $x$ for both EMC and  HERMES experiments is $x \sim0.02$).
We checked that by using the EKS98 parametrization \cite{EKS98} of PDF's
with shadowing corrections, the multiplicity ratios 
\eqref{multratioz} and \eqref{multrationu} change by less than 1\%.

Furthermore, we analyzed charged hadron production with both
the Khniel-Kramer-P\"otter parametrization of the FF at leading order 
(KKPLO)\cite{KKP} and the leading order Kretzer's parametrization
(KLO) \cite{K}. The different parametrization of the FF produces
a 1\% difference in the multiplicity ratios.
In the rest of the paper we shall use the KLO parametrization of the FF
since it gives charge-separated fragmentation functions for pions and
kaons, which are necessary for a comparison of the theoretical 
model with the HERMES data. 
Note, however, that by isospin symmetry arguments it is possible to relate
charge-averaged FF's, like KKPLO, to charge-separated ones at the cost of an
additional parameter \cite{ZFL02}. 

\begin{figure}[t]
\begin{center}
\begin{minipage}[t]{15.2cm}
\begin{center}
\epsfig{figure=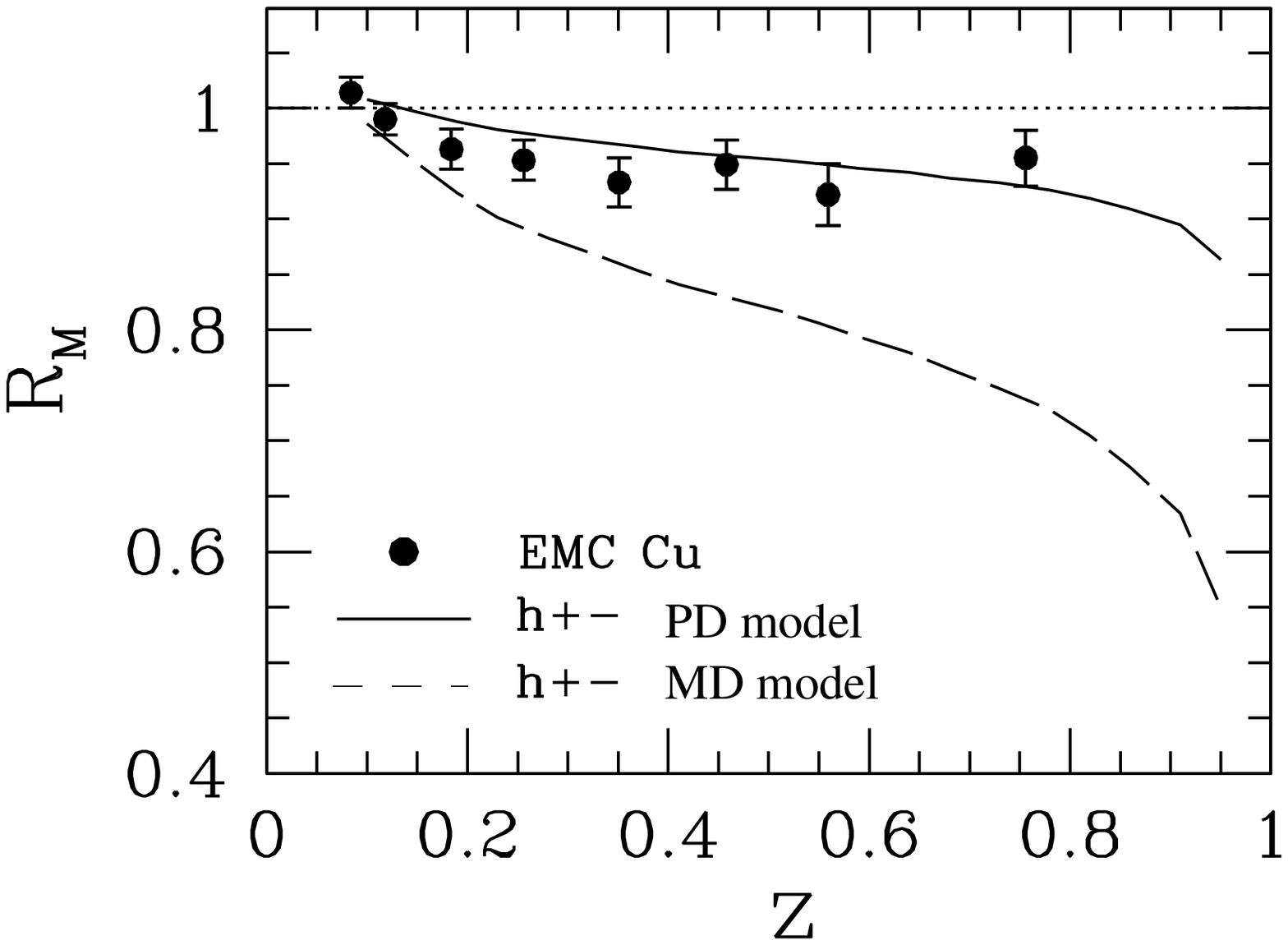,width=7.5cm
,clip=,bbllx=0pt,bblly=150pt,bburx=500pt,bbury=570pt}
\epsfig{figure=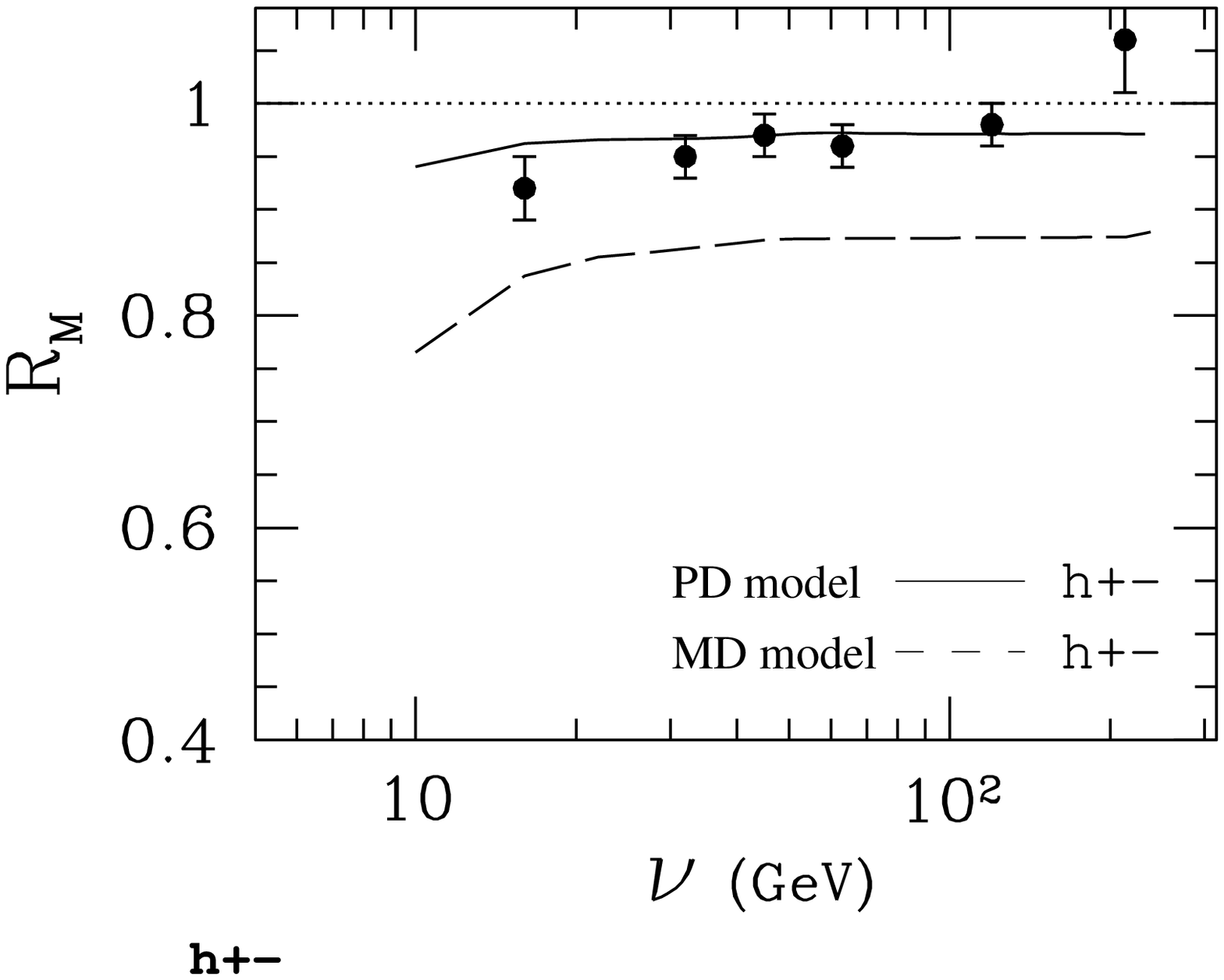,width=7.5cm
,clip=,bbllx=0pt,bblly=150pt,bburx=500pt,bbury=570pt}
\vskip-1.1cm\ 
\caption{\footnotesize
Theoretical multiplicity ratios $R_M$ as functions of $z$ and $\nu$ 
in the PD (solid lines) and MD
(dashed lines) models are compared with EMC data \cite{EMC}
on Cu. The theoretical curves between 10 GeV $<\nu<$
30 GeV are computed with the experimental acceptance corresponding
to a  beam energy of 100 GeV (see \appref{app:EMCandHERMES}).}
\label{fig:res1}
\end{center}
\end{minipage}
\end{center}
\vskip-.4cm
\end{figure}

Due to Lorentz dilatation, at large values of
$\nu$ the hadrons are expected to form mainly outside the nucleus, so
that the effect of reinteractions of the hadron with the nucleus are
minimal. For this reason, the EMC data on hadron production in
high energy DIS scattering on copper are ideal for a comparison of
the partial and the maximal
deconfinement model. The results for the $z$- and $\nu$-dependence of the
multiplicity ratio are shown in \figref{fig:res1} and
compared with EMC data \cite{EMC}. The maximal deconfinement model
(dashed lines) underestimates the data over nearly the whole 
range in $z$ and $\nu$. On the contrary the partial deconfinement model
(solid lines) better reproduces  both the $z$ and $\nu$   
experimental distributions
but overestimates the  data at low $\nu$ and $z$,  thus suggesting
that additional processes contribute in this region.  
As the maximal deconfinement model overpredicts the effect of
deconfinement in the nucleus, in the following we shall
consider only the  partial deconfinement model.

\section{Formation length distribution in the Lund model}
\label{sec:probdist}
\setcounter{equation}{0}

The results presented in  \figref{fig:res1} show that the partial deconfinement
rescaling model overestimates the multiplicity data slightly at small
energy transfers $\nu$ and at small $z$.
Therefore, we have to consider in addition the 
formation of the hadron and its subsequent interaction in 
the nuclear medium.
As the hadron formation length decreases for lower $\nu$,
the effect of
nuclear interaction becomes  more relevant in the kinematic region of 
the HERMES experiment, and the effect is amplified
in a heavy target as the formation length 
is comparable with the nuclear radius.
For a realistic calculation, a good understanding of the 
formation mechanism is necessary.

\begin{figure}[t]
\begin{center}
\begin{minipage}[t]{15.8cm}
\begin{center}
\parbox[c]{8.1cm}{\hspace*{-.7cm}
\vspace*{-1.cm}  
\epsfig{figure=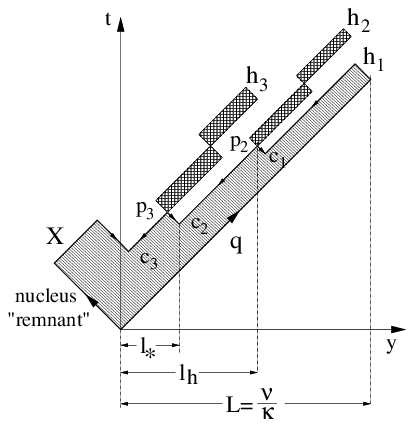,width=8cm}}
\parbox[c]{7cm}{
\caption{\footnotesize
Formation lengths of the prehadron and hadron are shown
in the Lund string model. The shaded
area represents the world-sheet swept by the string. Produced
hadrons, $h_i$, are ordered according to 
their rank $i$ (here $i=1,2,3$ and 3 hadrons are produced). 
At points $C_i$ prehadrons are created and at  
points $P_i$ the hadrons $h_i$ are formed. 
The formation lengths $l_*$ and $l_h$ at the
bottom of the figure refer to the second-rank hadron $h_2$.}
\label{fig:space-time}
}
\end{center}
\end{minipage}
\end{center}
\vskip.4cm
\end{figure}

In this paper we  follow the LUND model  for the fragmentation
process \cite{LundPhysRep,LundBook} and we apply the hadron formation and
interaction discussed in  \refref{BG87} to the DIS process.
A picture of the space-time development of the fragmentation process in 
the Lund model is presented in \figref{fig:space-time}. 
The process begins at t=0 and y=0 when the quark $q$ is
ejected from a nucleon by the virtual photon . 
The quark propagates in the positive longitudinal direction
$y$ and a colour string is formed between the quark and the target 
remnant.
The  string maps out the lightly shaded space time area.
It breaks at points $C_i$ into smaller pieces due to quark-antiquark
pair creation.
When a quark moving in the positive direction meets an antiquark
moving in the negative direction, they interact and form a so called
``yo-yo'' state, which is identified with a final state meson. This
happens at points $P_i$ in the figure, where the created hadron is
denoted by $h_i$. We shall denote by $z_i$ the fraction of the energy 
it carries. Hadrons are ordered according to their rank
$i$ \cite{LundPhysRep}. Note that the first-rank hadron is always
created at the end of the string, whose length is in our case
\begin{align*}
    L=\frac{\nu}{\kappa_A} \ ,
\end{align*}
where $\kappa_A$ is the string tension in the nuclear environment.
If  partial  deconfinement occurs, 
also the string tension is rescaled  in the nucleus:
\begin{align}
    \kappa_A \lambda_A^2 = \kappa \lambda_0^2 \ ,
\end{align}
where $\kappa \sim 1$ GeV/fm is the string tension in the vacuum.
Since the string tension is the physical quantity that 
sets the confinement scale, a larger confinement
scale corresponds to a smaller string tension.

\begin{figure}[t]
\begin{center}
\vspace*{-5.cm}
\begin{minipage}[t]{15cm}
\begin{center}
\parbox[c]{7.7cm}
{
\hspace*{-4.cm}\epsfig{figure=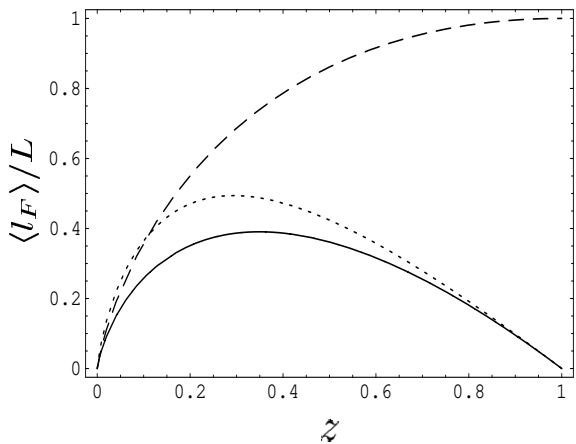,width=14.5cm, height=25.cm}}
\parbox[c]{6.8cm}{
\vspace*{-7.5cm} 
\caption{\footnotesize The average normalized prehadron formation length
  $\vev{l_*}/L$ computed in standard Lund model 
  \eqeqref{Lundaverlf} with $C$=0.3 (solid line), compared with
  the one calculated from \eqeqref{EMCaverlf} used in the EMC
  analysis (dotted line). The normalized hadron  (or yo-yo)
  formation length  $\vev{l_h}/L$ is shown as a dashed curve. }  
\label{fig:averagelf}
}
\end{center}
\end{minipage}
\end{center}
\vskip-14cm
\end{figure}

Let us focus our attention on the $h_2$ hadron in 
\figref{fig:space-time}. There are two relevant lengths for the
fragmentation process: 
\begin{itemize}
\item[{\it (i)}]
prehadron (or constituent) formation length $l_*$ at which the first 
constituent of $h_2$ is created. 
\item[{\it (ii)}] 
the hadron (or yo-yo) formation length $l_h\leq L $ at which the hadron
$h_2$ is formed.
\end{itemize}
In the Lund model these two lengths are related in the following way
\cite{LundBook,BG87}:
\begin{align}
    l_h = l_* + z_2 L \ ,
  \label{l*vslh}
\end{align}
At fixed $z$ they both increase linearly with the
virtual photon energy $\nu$. However, as functions of $z$ they 
behave rather differently, especially at $z \ra 1 $, where $l_*\ra 0$ 
and $l_h\ra L$ (see \figref{fig:averagelf}). 
Therefore, we have two distinct  scenarios
of fragmentation, whether the prehadronic state interacts with
the nucleus, or not. 
Analysis of the data on hadron-nucleus and lepton-nucleus collisions
strongly suggest that the prehadron  behaves as an hadron-like object
long before final state hadrons actually appear \cite{BG87,BC89}. 
Also an analysis of EMC and SLAC data confirms the importance of the
prehadron interactions \cite{C89}.

In \appref{app:Lundcomps}
we derive the probability distribution ${\cal P}_*$  that a
prehadronic state is formed at a formation length $y$ with momentum 
fraction $z$. In the case that the prehadron is composed of
$q=u,d,s$ flavours, the distribution reads
\begin{align}
    {\cal P}_*(y;z,L) \ = \ &
        \frac{zL}{y-zL}
        \left[ \frac{y}{(y+zL)(1-z)} \right]^{C} 
        \nonumber \\
    & \times \left\{ \delta[y-(1-z)L] 
          + \frac{1+C}{y-zL} \theta[(1-z)L-y]\right\} \theta[y] \ ,
  \label{Lundprobdist} 
\end{align}
where $C\approx 0.3$ is the only parameter entering this expression. 
We can then compute the hadron average
formation length $\vev{l_F}$, which we identify 
with the prehadronic formation
length, $\vev{l_F}=\vev{l_*}(z,L) = \int dy \, y\, {\cal P}_*(y;z,L)$:
\begin{align}
  \vev{l_F} = \left[ 1 + \frac{1+C}{2+C}\, \frac{1-z}{z^{2+C}} \,
       {}_2F_1\Big( 2+C,2+C;3+C;\frac{z-1}{z} \Big)
       \right] (1-z)\,z\,L \ ,
  \label{Lundaverlf}
\end{align}
where ${}_2F_1$ is the Gauss' hypergeometric function \cite{Magnus}.
Note that as $z\ra1$ the average formation length 
$\vev{l_F} \ra (1-z)\frac{\nu}{\kappa}$   
behaves similarly to the formation length
suggested by the gluon bremsstrahlung model of \refref{KNP96}. 
Indeed, quantum mechanics yields a large  
energy for an intermediate state consisting of a quark and a gluon
with $z$ and $1-z$ momentum fractions when $z \ra 1$. 
Consequently, if the hadron takes almost all the energy ($z\ra 1$)
then the quark-gluon system is short lived and must hadronize almost 
immediately \cite{KOPNY}.

In  \figref{fig:averagelf} we show the average 
formation length computed with \eqeqref{Lundaverlf} and  $C=0.3$
as in  the standard Lund model (solid line). 
This is the formula we use in the next sections.
For comparison (dotted line) 
we show also the average formation length 
used in the analysis of the EMC data \cite{PAVEL}, 
\begin{align} 
  \vev{l_F} = \left[ \frac{\ln(1/z^2)-1+z^2}{1-z^2} \right] z L \ ,
 \label{EMCaverlf}
\end{align}
which overestimates up to 30\% the Lund model result. The above
formation length is obtained in the Lund model with a non-standard
choice of parameters, as described in \appref{app:Lundcomps}.
Finally, the contribution of the hadron (or yo-yo) formation lenght
$l_h$ in the standard Lund model is shown with a dashed curve.

\section{Absorption of the produced hadron in the nucleus}
\label{sec:nuclearabsorption}
\setcounter{equation}{0}

\begin{figure}[t]
\begin{center}
\begin{minipage}[t]{15cm}
\begin{center}
\parbox[c]{7.7cm}{\epsfig{figure=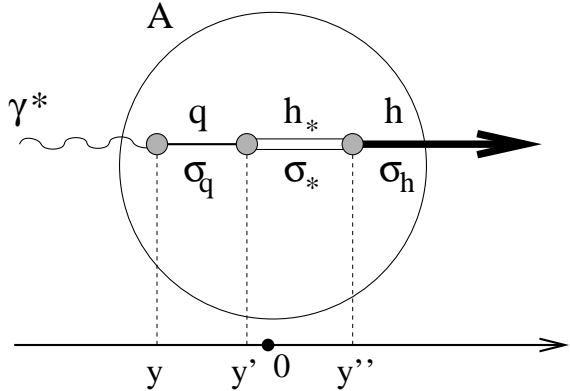,width=7.5cm}}
\parbox[c]{6.8cm}{
\caption{\footnotesize
Nuclear attenuation of a hadron $h$: the virtual photon
$\gamma^*$ interacts with a quark $q$ at a longitudinal coordinate $y$;
the quark turns into a ``prehadronic''
state $h_*$ at position $y\,'$ and the hadron $h$ is formed at $y\,''$.  
Each state interacts with the surrounding nucleons with a
cross-section $\sigma_q$, $\sigma_*$ and $\sigma_h$, respectively. }
\label{fig:nuc-abs}
}
\end{center}
\end{minipage}
\end{center}
\vskip-.4cm
\end{figure}

The spatial evolution of a quark $q$ created by the virtual photon 
$\gamma^*$ at a longitudinal position $y$ is shown in  \figref{fig:nuc-abs}
in all intermediate stages. In the first stage the quark propagates
to the position $y\,'>y$ where a prehadronic state $h_*$ 
is formed. In the second stage the final hadron $h$ is
created at the point $y\,''>y\,'$. 

The quark, the prehadronic state and the
final hadron propagate through  nuclear matter and 
interact with the surrounding nucleons. As a result,
each of the three states may undergo  inelastic
interactions and/or lose longitudinal energy through 
elastic rescatterings. 
In a first approximation we may assume that 
the final hadron with a value of $z$ 
corresponding to the fragmentation process in vacuum 
will be observed provided that none
of the three propagating states has interacted with the  nucleus.
By defining the nuclear absorption factor ${\cal N}_A(z,\nu)$ 
as the probability that neither the $q$,
$h_*$, nor $h$ have interacted with a nucleon,
the multiplicity ratios of 
Eqs. \eqref{dnhdz} and \eqref{dnhdnu} are modified
as follows:
\begin{align}
    \frac{1}{N_A^{DIS}}\frac{dN_A^h}{dz} & = \frac{1}{\sigma^{\ell A}} 
         \int_{\rm exp.\ cuts}\hspace*{-1.0cm}
         dx\, d\nu \sum_f e^2_f\, q_f(x,\xi_A\,Q^2) \,
         \frac{d\sigma^{\ell q}}{dx\,d\nu} \, D_f^{h}(z,\xi_A\,Q^2)  
         \, {\cal N}_A(z,\nu) \ ,
 \label{dnhdzabs} \\
    \frac{1}{N_A^{DIS}}\frac{dN_A^h}{d\nu} & = \frac{1}{\sigma^{\ell A}} 
         \int_{\rm exp.\ cuts}\hspace*{-1.0cm}
         dx\, dz \sum_f e^2_f\, q_f(x,\xi_A\,Q^2) \,
         \frac{d\sigma^{\ell q}}{dx\,d\nu} \, D_f^{h}(z,\xi_A\,Q^2)  
         \, {\cal N}_A(z,\nu) \ .
 \label{dnhdnuabs} 
\end{align}
While the fragmentation functions $D_f^{h}$ are sensitive to the
virtuality $Q^2$ of the fragmentation process in the medium, the
nuclear absorption factor ${\cal N}_A$, which depends on $z$ and $\nu$, 
is sensitive to the hadron energy in the rest frame of the
nucleus. 

We will  investigate two models for the
computation of the nuclear absorption factor: the Bialas-Gyulassy
(BG) model, \cite{BG87}, and the Bialas-Chmaj (BC) model, \cite{BC83}.
These models require
the probability distribution function of
the hadron formation length and the average formation
length  discussed and derived in \secref{sec:probdist}.
In  order to reduce the number of the  parameters of these models 
we set the quark-nucleon cross-section $\sigma_q=0$,
since the analysis of the DIS data on $R_M^h$ reported in
\cite{EMC,PAVEL,HERMES} 
does not indicate a significant interaction of the quark with the nuclear
medium. Moreover, we  do not give the formulae for a separate
evolution 
of the prehadron to the full hadronic size in the main text. Then, 
the prehadronic state $h_*$ and the hadron $h$ are treated
effectively as a
single object $h_*=h$ 
created at $y\,'$ which propagates in the
nucleus with a single cross-section $\sigma_*=\sigma_h$. 
The effect of distinguishing the prehadronic state by a different 
cross section
$\sigma_* \neq \sigma_h$ is  discussed in \secref{sec:chargedhadrons}
and \ref{sec:flavours}, and the formulae needed in the computations of
the absorption effect are discussed in App.C.

\subsection{Bialas-Gyulassy (BG) model}

In the Bialas-Gyulassy (BG) model \cite{BG87} the nuclear absorption 
for a one-step
fragmentation process with
$\sigma_*=\sigma_h$ depends on three parts:
the nuclear density profile $\rho_A(b,y)$ normalized to unity 
as a function of impact parameter $b$
and longitudinal coordinate $y$; the probability distribution 
${\cal P}_*(y\,'-y,z,L)$ that the prehadron is
formed after the length  $y\,'-y$;
and the probability $[W_0(y\,')]^{A-1}$ that the hadron created at $y\,'$ 
does not interact with the nucleus.
The nuclear absorption factor has the following form:
\begin{align}
    {\cal N}_A = \int d^2b \int_{-\infty}^{\infty} dy \, 
        \rho_A(b,y) \int_{y}^{\infty} dy\,'  
        {\cal P}_*(y\,'-y,z,L) \big[ W_0(y\,') \big]^{A-1} \ ,
 \label{BG1step}
\end{align}
where
\begin{align}
    W_0(y\,') = 1 
        - \sigma_h \int_{y\,'}^\infty \hspace*{-.2cm}dy_*\,
        \rho_A(b,y_*) \ .
 \label{BG-W0}
\end{align}
For ${\cal P}_*$ we will use the probability distribution
\eqref{Lundprobdist}
discussed in \secref{sec:probdist}.

\subsection{Bialas-Chmaj (BC) model}

The Bialas-Chmaj (BC) model \cite{BC83} 
is a simpler version of the BG-model 
for  including nuclear absorption.
In the BC model the nuclear absorption factor 
depends on the probability $ P_h(y\,',y)$ that the hadron $h_*=h$ 
is formed at a distance $y\,'-y$ from the $\gamma^*q$ interaction
point, 
\begin{align}
    P_h(y\,'-y) = 1 - e^{\,-(y\,'-y)/\vev{l_F}} \ ,
  \label{PqPh}
\end{align}
and on the  survival probability $S_A(b,y)^{A-1}$  of this hadron, 
where for a one step fragmentation process with $\sigma_*=\sigma_h$ 
one has
\begin{align}
    S_A(b,y) = 1 
        - \sigma_h \int_y^\infty \hspace*{-.2cm}dy\,'\,
        P_h(y\,'-y)  \rho_A(b,y\,')\ .
 \label{BCabs}
\end{align}
The nuclear absorption factor is then written as:
\begin{align*}
    {\cal N}_A = \int d^2b \int_{-\infty}^{\infty} 
        dy \, \rho_A(b,y)\big[ S_A(b,y) \big]^{A-1} \ ,
\end{align*}
where  the average formation length
$\vev{l_F}=\vev{l_F}(z,\nu)$ we will use has been derived in
\secref{sec:probdist}, \eqeqref{Lundaverlf}.

To understand the BC model and its relationship to the   
BG model, \eqeqref{BG1step}, we can approximate $\rho_A(y,b)$ with a
uniform nuclear density $\rho_A(y,b)=\rho_{0}/A$ inside a hard sphere.
Here $\rho_{0}$ is the
nuclear matter density. In this approximation the 
integrations in \eqeqref{BCabs} may be carried out explicitly 
and yield
\begin{align}
   {\cal N}_A^{(BC)} = \left[  
     1-\frac{\sigma_h \rho_0}{A} 
     \left\{ R(b)-\left(y-\vev{l_F^{\rm eff}}\right) 
     \theta\left(R(b)-y\right)\theta\left( R-b \right) \right\}
     \right]^{A-1} \ .
  \label{BCanalytic}
\end{align}
Here $R(b)^2=\sqrt{R^2-b^2}$ and 
the effective  formation length
$\vev{l_F^{\rm eff}}$ is defined as:
\begin{align}
  \vev{l_F^{\rm eff}} = \vev{l_F}
     \left[ 1-\esp{-\left( R(b)-y \right) / \vev{l_F}} \right] \ .
  \label{leff}
\end{align} 
It is then easy to see that \eqeqref{BCanalytic} can be derived from
the Bialas-Gyulassy formula \eqref{BG1step} if we choose
\[
   {\cal P}_*(y\,'-y) = 
     \delta\left[ (y\,'-y) - \vev{l_F^{\rm eff}} \right] \ .
\] 
In this sense, the BC model for nuclear absorption is an effective
model for absorption which describes the fragmentation process
by modifying the average formation length according to \eqeqref{leff},
and letting the hadron to be formed suddenly after a well defined 
formation length,
instead of assuming a probability distribution in formation
lengths. Nonetheless, as we shall discuss
in \secref{sec:chargedhadrons} and \secref{sec:flavours}, the BC model
gives a useful phenomenological description of nuclear absorption
processes.

\section{Numerical results for multiplicity ratios in nuclei}
\label{sec:chargedhadrons}
\setcounter{equation}{0}

In this section we discuss the results of our model for
semi-inclusive charged hadron production in nuclei including effects of 
perturbative gluon radiation and absorption, and compare them 
to EMC and HERMES experimental data.
 
\begin{figure}[t]
\begin{center}
\begin{minipage}[t]{16cm}
\begin{center}
\parbox{9cm}{
\epsfig{figure=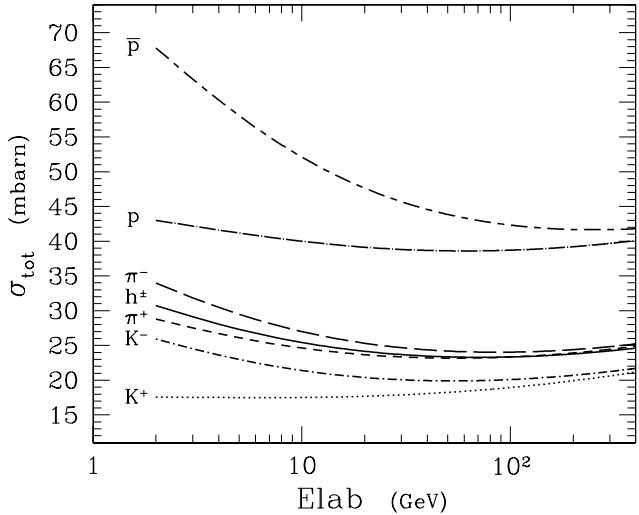,width=9cm
,clip=,bbllx=30pt,bblly=250pt,bburx=600pt,bbury=710pt}
}
\parbox{6cm}{
\caption{\footnotesize
Isospin averaged 
hadron-nucleon total cross section versus the hadron energy
$E_h=E_{lab}$, in the case of krypton nucleus, 
for positive and negative pions and kaons, and protons and antiprotons.
The charged hadron cross-section $ \sigma_h$ 
has been computed as weighted sum of individual cross sections for
pions, kaons, protons and antiprotons.}
\label{fig:totxsec} }
\end{center}
\end{minipage}
\end{center}
\vskip-1cm
\end{figure}

In the computation of the nuclear absorption factor, we use as a
density for the deuteron the sum of the 
Reid's soft-core S- and D-wave functions squared \cite{ReidSC}. 
For heavier nuclei we use a Woods-Saxon density with radius
$R_A=1.12 A^{1/3} - 0.86 A^{-1/3}$ fm.

As discussed in \secref{sec:nuclearabsorption},
the fragmentation of the quark is considered without differentiation
between the prehadron-nucleon and hadron-nucleon cross sections.
For the hadron-nucleon cross sections, we take the isospin averaged 
hadron-proton  $\sigma_{hp}$
and hadron-neutron $\sigma_{hn}$ total  cross sections:
\begin{align}
    \sigma_h(E_h) = \frac{1}{A} \big[ Z \sigma_{hp}(E_h) + (A-Z)
    \sigma_{hn}(E_h) \big]. \
\end{align}
As the total hadron-nucleon cross sections include also the elastic
ones, the corresponding results  
represent an upper limit with respect to the nuclear absorption effect
in the multiplicity ratios. The energy
dependence of $\sigma_{hp}$ and $\sigma_{hn}$ \cite{PDB} has been 
explicitly  taken into account, and is at variance with respect to
the original models  \cite{BG87,BC83}.
The energy behavior of the total hadron-nucleon cross section is 
shown in \figref{fig:totxsec},
in the case of a krypton
target,  for positive/negative pions and kaons. 
The corresponding charged hadron cross section has been derived as weighted
sum of individual cross sections for pions, kaons, protons and antiprotons.
The weighted sum has been performed by taking into account the measured
relative yield of various hadrons reported by the HERMES experiment in the 3-23
GeV region \cite{HERMES} as this region is very sensitive to the variation of
the hadron-nucleon cross section. 

\begin{figure}[t]
\begin{center}
\begin{minipage}[t]{16cm}
\begin{center}
\epsfig{figure=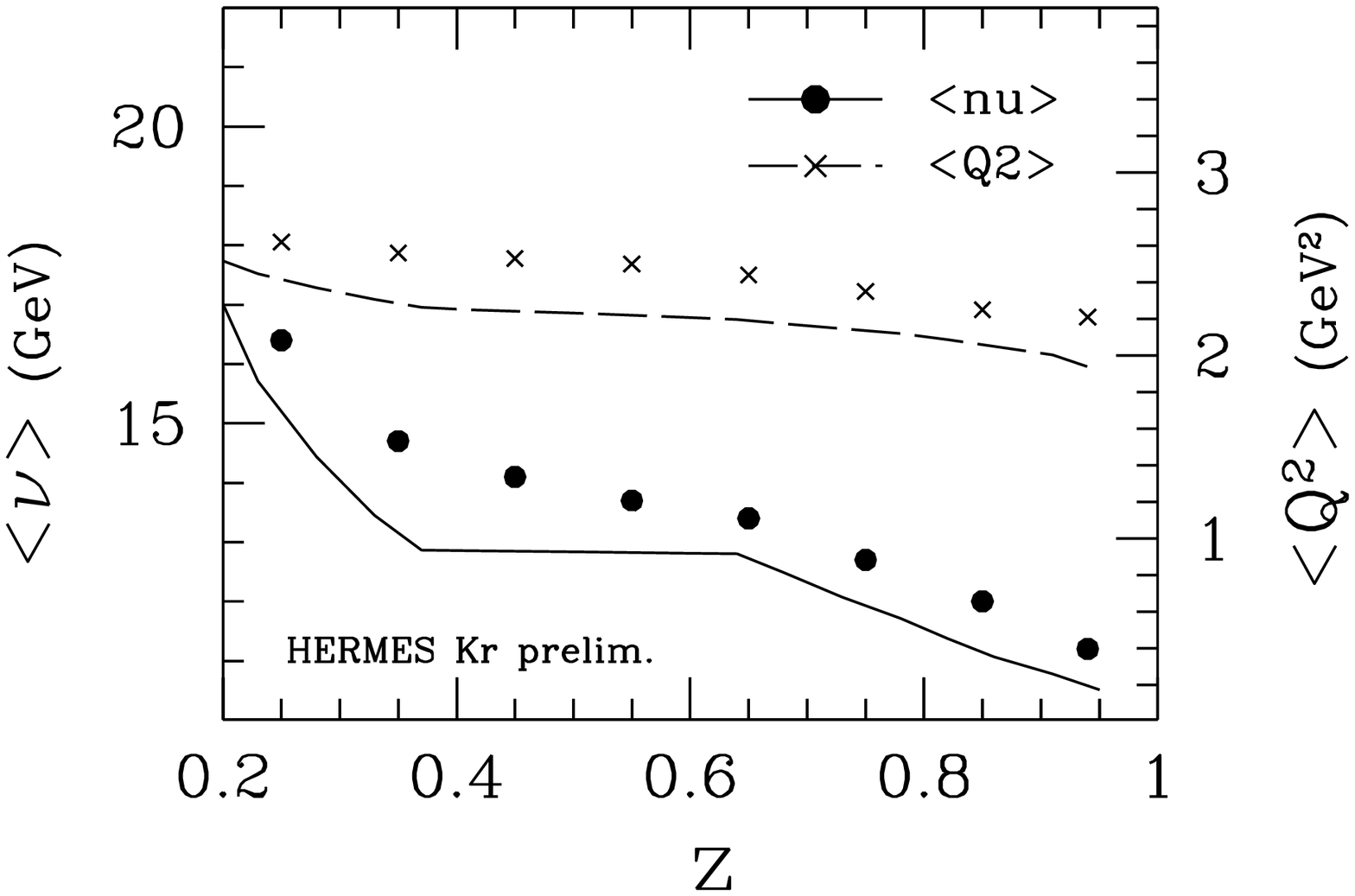,width=7.8cm
,clip=,bbllx=30pt,bblly=320pt,bburx=580pt,bbury=718pt}
\epsfig{figure=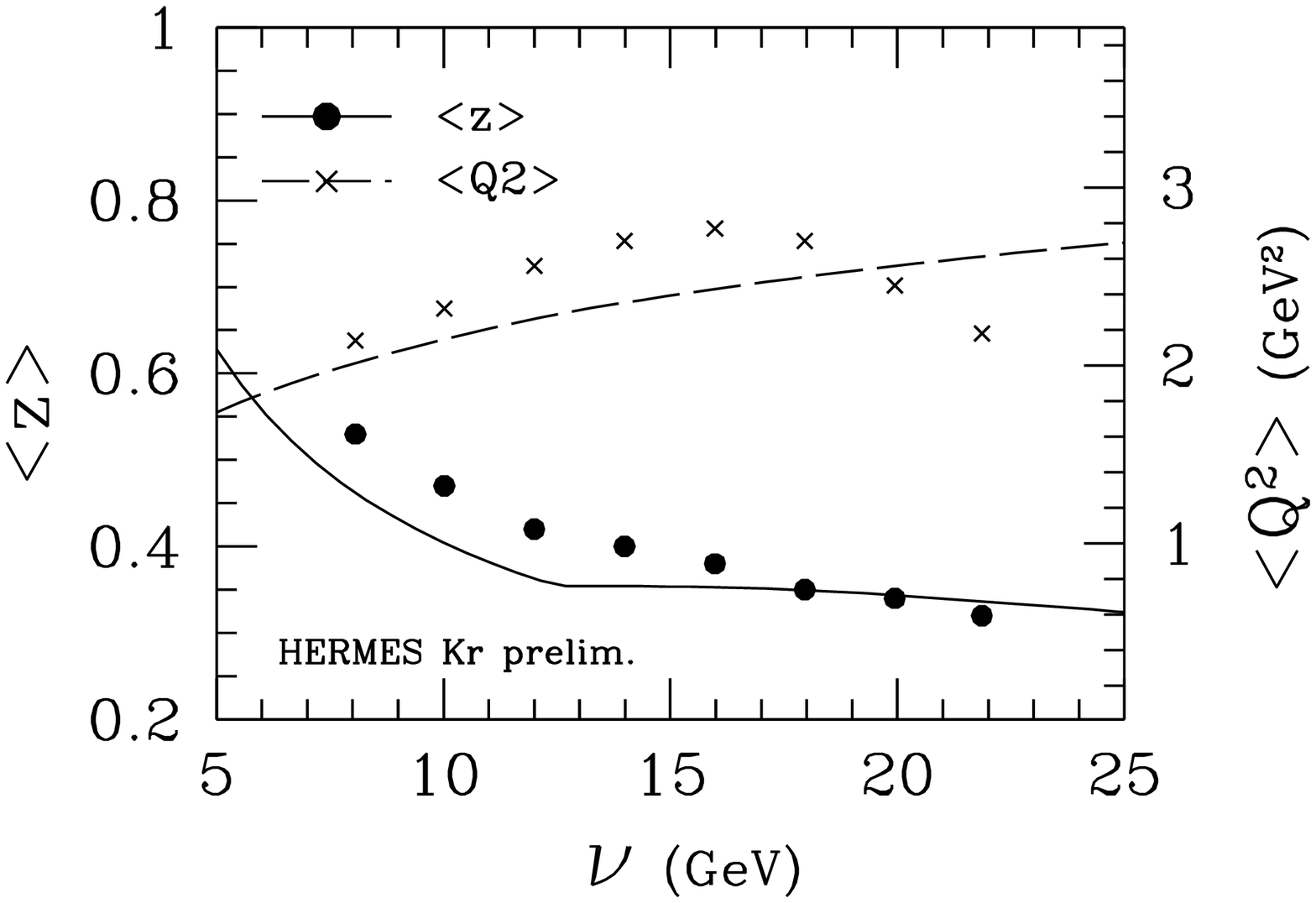,width=7.8cm
,clip=,bbllx=30pt,bblly=320pt,bburx=580pt,bbury=718pt}
\caption{\footnotesize
Average theoretical values of kinematic variables computed with the
LO formulae \eqref{dnhdz} and \eqref{dnhdnu} are compared with the
experimental kinematic variables at HERMES
for $\pi$ and $K$ production on Kr \cite{DATAHERMES}.
{\it Left:} $\vev{\nu}$ and  $\vev{Q^2}$ as a function of $z$.
{\it Right:} $\vev{z}$ and  $\vev{Q^2}$ as a function of $\nu$. }
  \label{fig:expave}
\end{center}
\end{minipage}
\end{center}
\vskip-.4cm
\end{figure}

As the model predictions are sensitive to the average kinematic 
variables used in the computations,
in \figref{fig:expave} we compare
our LO theoretical computations 
of the average $\nu$, $z$ and $Q^2$ with the 
corresponding experimental  values.
The model reproduces reasonably well the experimental $\langle Q^2 \rangle$,
$\langle z \rangle$ and $\langle \nu \rangle$ behaviour. 
The step structures found in the LO computation of 
$\langle \nu \rangle (z)$ and $\langle z \rangle (\nu)$ are due to the
experimental kinematic cuts shown in \figref{fig:kincuts} of
\appref{app:EMCandHERMES}.
The differences compared with the experimentally observed
smoother dependence of   $\langle \nu \rangle$ and  $\langle z
\rangle$ may come from NLO corrections or higher-twist terms.
Therefore,
in the computation of the multiplicity ratios we
replaced the arguments of \eqeqref{dnhdzabs}
by experimental values when available:
\begin{align*}
    q_f(x,\xi_A\,Q^2) 
        & \ \ \lora \ \ q_f(x,\xi_A\,\vev{Q^2}_{\rm exp}(z)) \\
    D_f^{h}(z,\xi_A\,Q^2) 
        & \ \ \lora \ \ D_f^{h}(z,\xi_A\,\vev{Q^2}_{\rm exp}(z)) \\
    {\cal N}_A(z,\nu) 
        & \ \ \lora \ \ {\cal N}_A(z,\vev{\nu}_{\rm exp}(z)) \ ,
\end{align*} 
and in \eqeqref{dnhdnuabs}
\begin{align*}
    q_f(x,\xi_A\,Q^2) 
        & \ \ \lora \ \ q_f(x,\xi_A\,\vev{Q^2}_{\rm exp}(\nu)) \\
    D_f^{h}(z,\xi_A\,Q^2) 
        & \ \ \lora \ \ D_f^{h}(z,\xi_A\,\vev{Q^2}_{\rm exp}(\nu)) \\
    {\cal N}_A(z,\nu) 
        & \ \ \lora \ \ {\cal N}_A(\vev{z}_{\rm exp}(\nu),\nu) \ .
\end{align*} 
The effect of using experimental averages instead of the LO-computed ones
by \eqseqref{dnhdzabs} and \eqeqref{dnhdnuabs}, maximally yields 2-3\% variation in the presented multiplicity ratios.
The multiplicity ratios for positive and negative charged hadrons are
calculated by using the fragmentation functions for $h^\pm$ 
in Kretzer's parameterization, with the meson formation time in the
Lund model, \eqseqref{Lundprobdist} and \eqref{Lundaverlf},
and with the average cross-section shown in \figref{fig:totxsec}. 
Since charged hadrons are dominated by meson production, this is a good
approximation to the sum of individual hadron fragmentation yields multiplied
by individual absorption factors. 
In \secref{sec:flavours} we will also calculate multiplicity ratios for pions
and kaons separately.
We leave out an extra calculation of baryon and antibaryon
production, since such a calculation would  require to know
the contribution of target fragmentation and to extract 
separate fragmentation functions for $p$ and
$\bar{p}$  from the charge averaged KKP parameterization.
Furthermore a different formation mechanism 
is possible for baryons or antibaryons,  
see \eqseqref{probdistgeneral} and \eqref{averlfgeneral}.

\begin{figure}[tb]
\begin{center}
\begin{minipage}[t]{15.2cm}
\begin{center}
\epsfig{figure=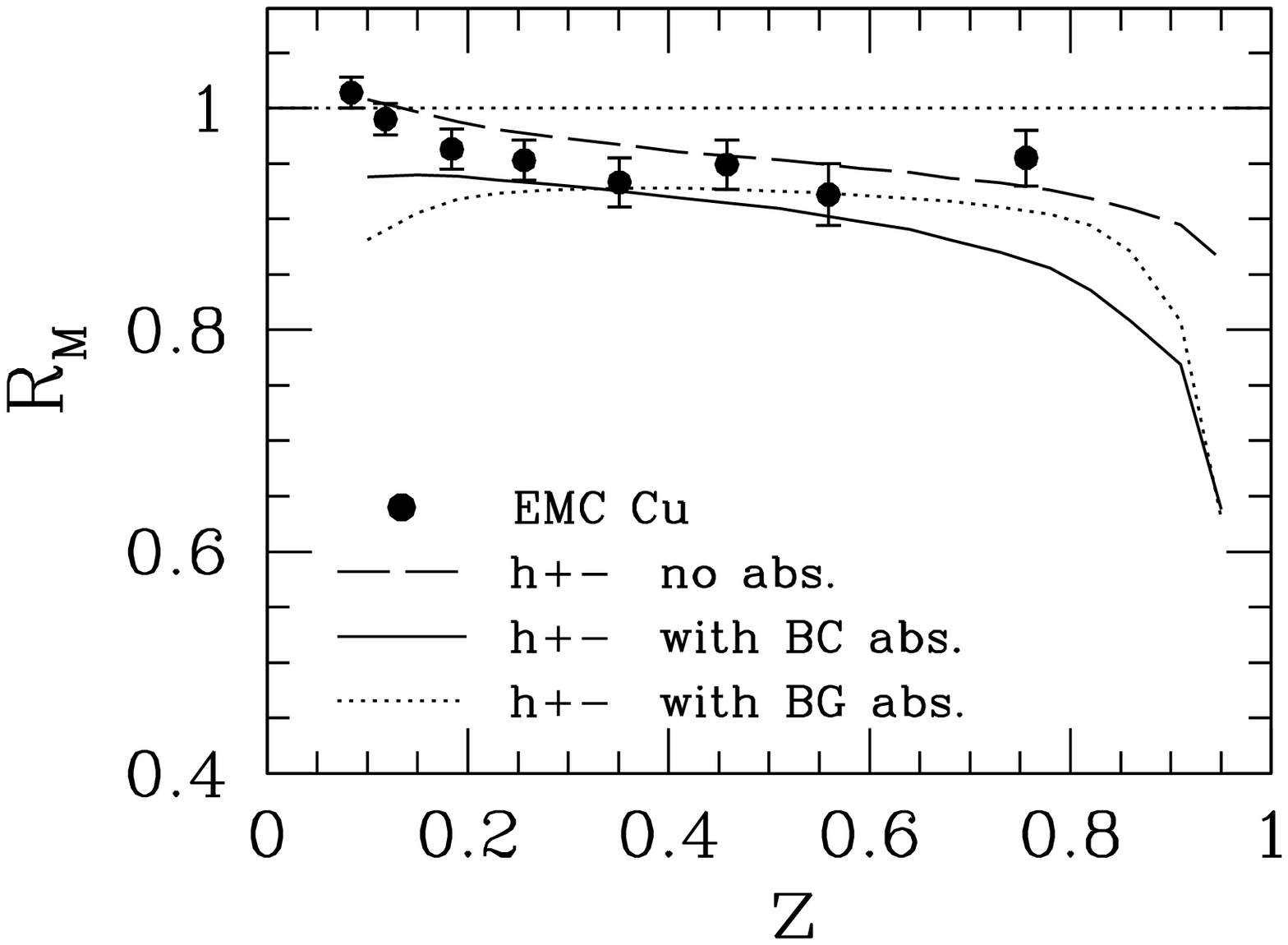,width=7.5cm
,clip=,bbllx=27pt,bblly=320pt,bburx=510pt,bbury=680pt}
\epsfig{figure=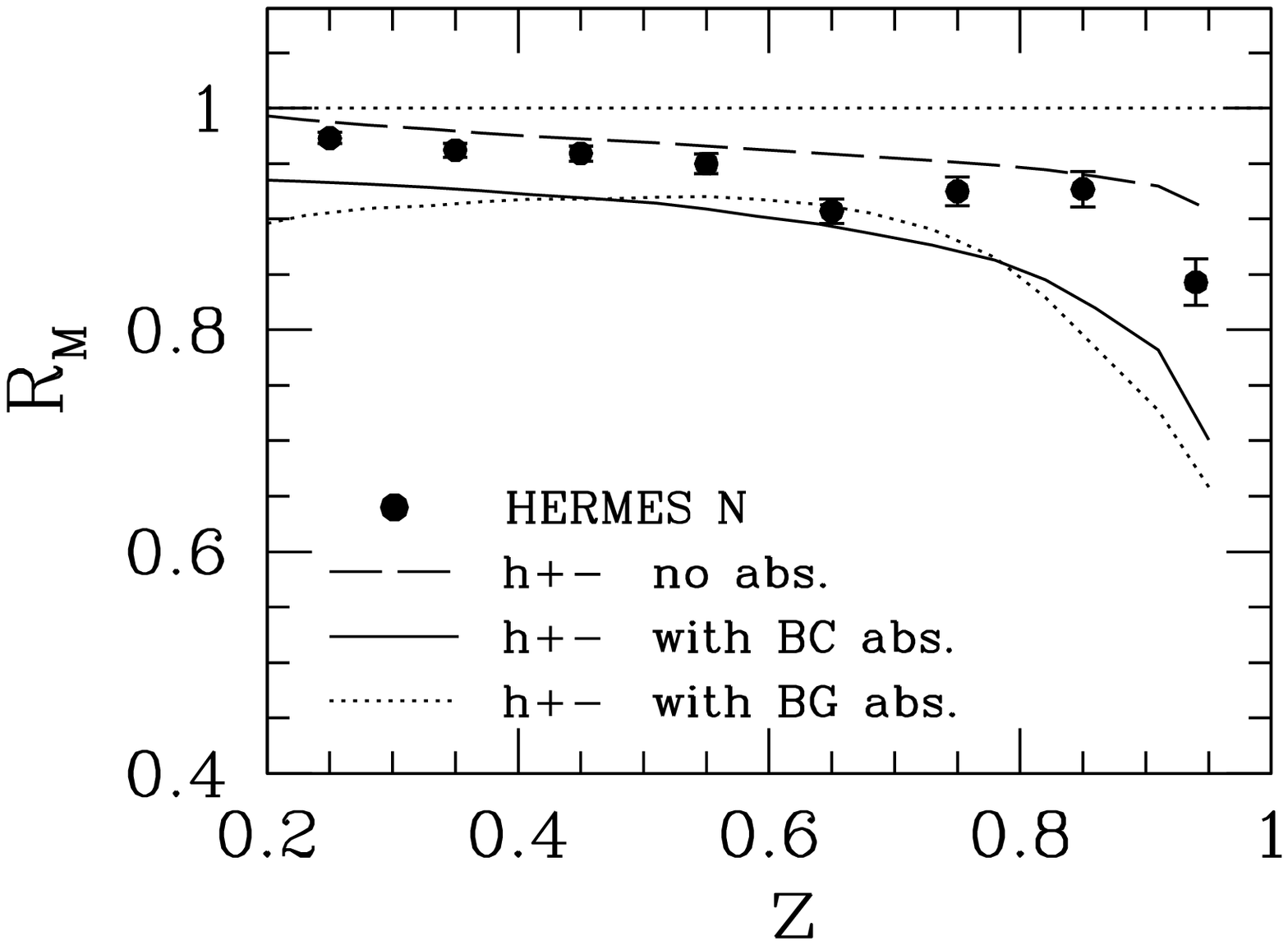,width=7.5cm
,clip=,bbllx=27pt,bblly=320pt,bburx=510pt,bbury=680pt}
\\
\epsfig{figure=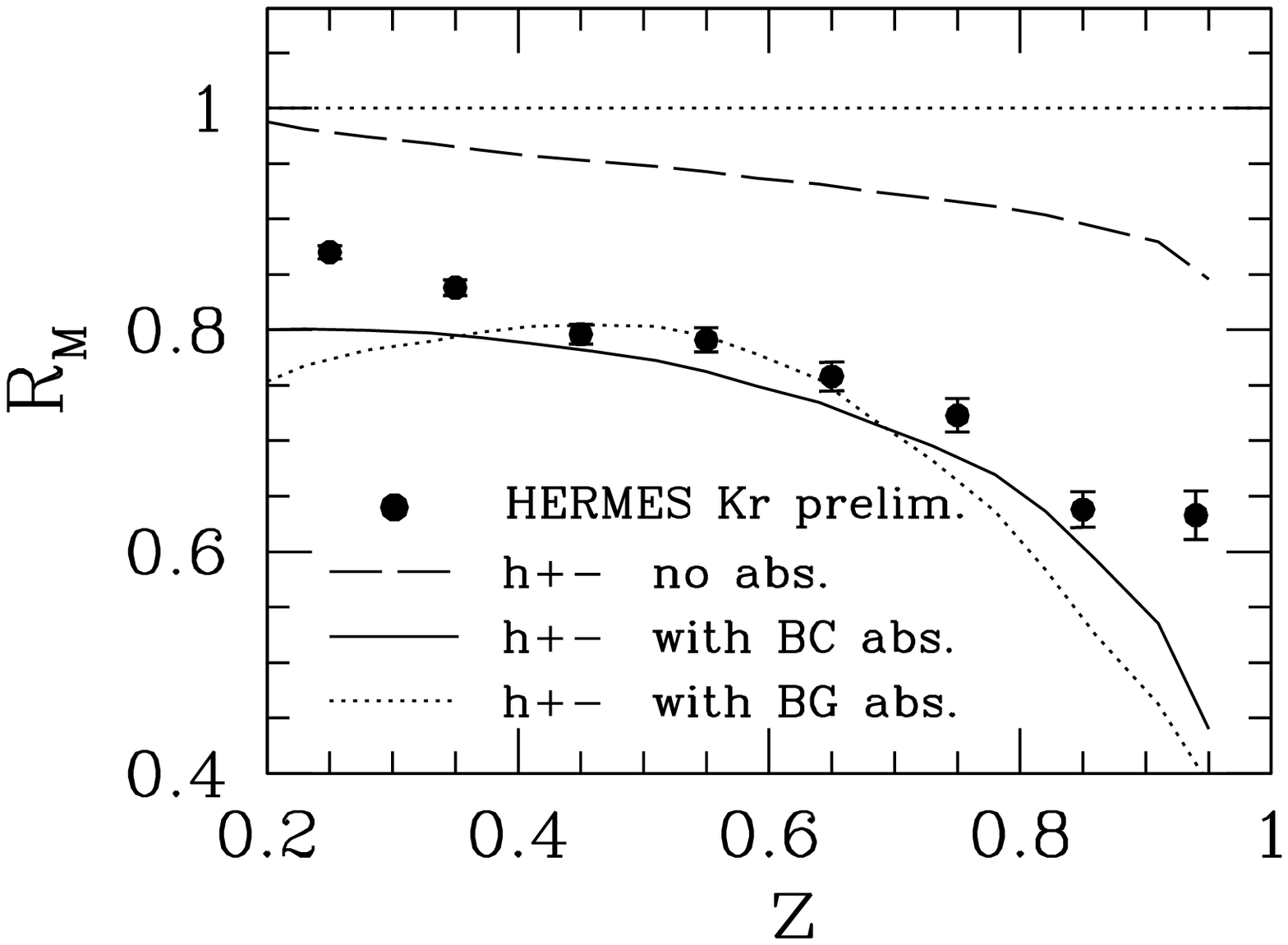,width=7.5cm
,clip=,bbllx=27pt,bblly=320pt,bburx=510pt,bbury=680pt}
\epsfig{figure=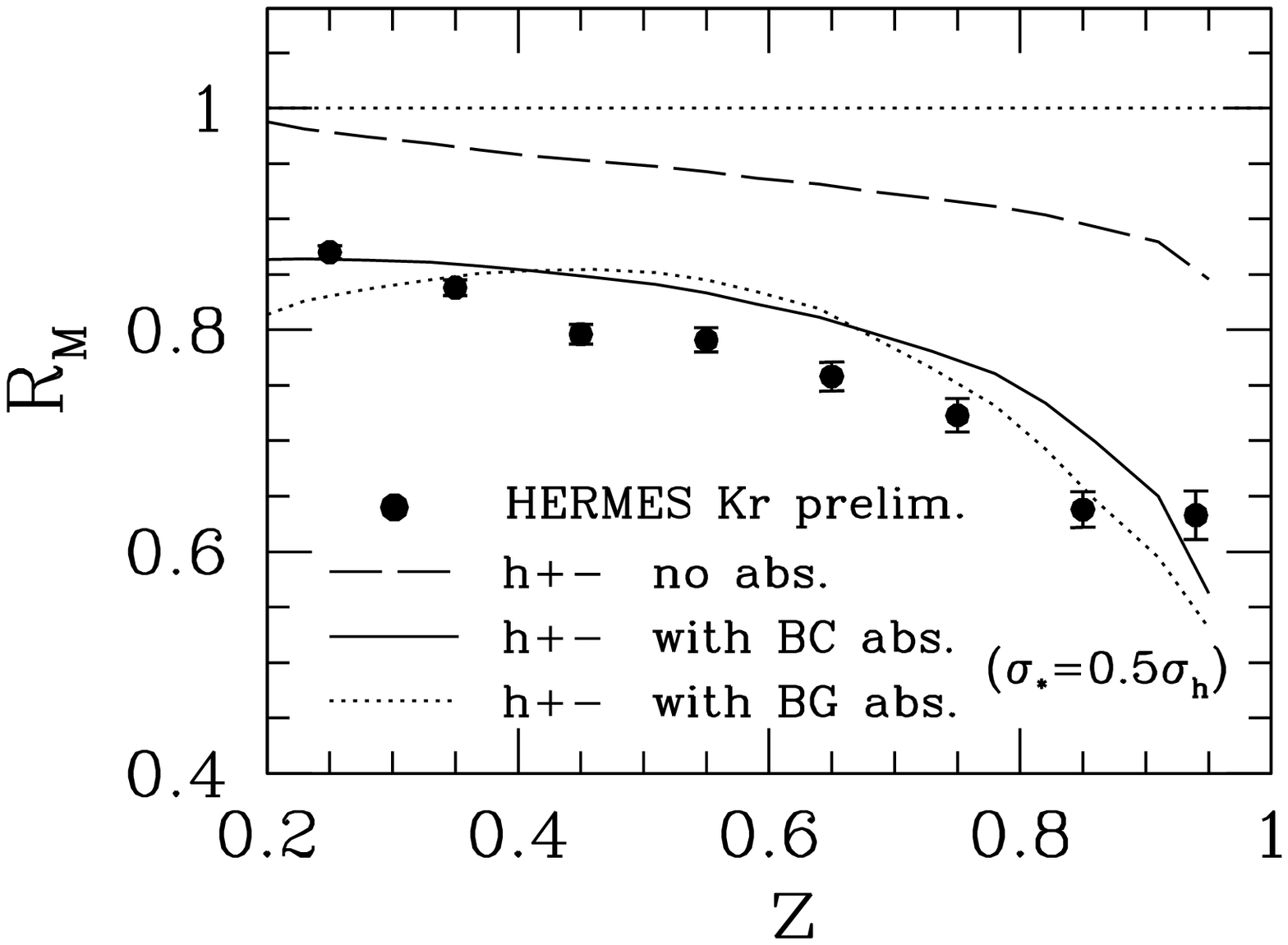,width=7.5cm
,clip=,bbllx=27pt,bblly=320pt,bburx=510pt,bbury=680pt}
\caption{\footnotesize
Theoretical multiplicity ratios of charged hadrons for Cu (EMC), 
Kr and N (HERMES) targets as function of $z$ compared 
with the data \cite{EMC,HERMES}.
Dashed lines show the predictions without absorption and only
rescaling, according to \eqseqref{dnhdz}-\eqseqref{DISxsec}; solid lines 
give the calculation with rescaling and 
absorption according to \eqseqref{BCabs} and \eqref{Lundaverlf} 
for the BC model; dotted lines with
rescaling plus absorption according to \eqeqref{BG1step} 
and \eqref{Lundprobdist} for the BG model. 
In the bottom right plot, the solid and dotted curves represent the
result of the 
computations with $\sigma_*=0.5 \, \sigma_h$.
The error bars represent the statistical
uncertainty only. 
Systematic uncertainties  for the HERMES nitrogen and krypton data are
3\% and 3.5\% respectively \cite{HERMES,DATAHERMES}.
}
 \label{fig:absresc}
\end{center}
\end{minipage}
\end{center}
\vskip-.4cm
\end{figure}

The theoretical calculations of the multiplicity ratio for 
charged hadrons are presented 
in \figref{fig:absresc} and 
 compared with the EMC and HERMES 
data for different nuclei.
In the figure, the three different  theoretical curves  correspond to
calculations with rescaled fragmentation functions alone (dashed curve),
with rescaling plus nuclear absorption in the BC 
model (solid curve), and with rescaling plus absorption
in the BG model (dotted curve).

In the case of BG absorption with the probability distribution 
of \eqeqref{Lundprobdist}, the general shape of the curve deviates 
from the experimental behavior and the calculation underestimates the
multiplicity ratio both at small and large $z$.  
In the case of BC absorption there is a better agreement with the
experimental behaviour. Note that in this case we used the average
formation length of \eqeqref{Lundaverlf} where we fixed the effective string
tension $\kappa_{BC} = 0.4$ GeV/fm in order to agree with the
experimental data. 
As already observed in \refsref{EMC,PAVEL} and \cite{AP02} the Bialas-Chmaj 
model with $\kappa=1$ GeV/fm overestimates by far the 
absorption effects, while, with a reduced effective string tension,
a reasonably good description of the data is obtained inside the
statistical and systematic errors.
 
The observed differences between the data and the model predictions in
the low-$z$ region  can be 
interpreted in several ways. In the  described models for hadron
absorption, a hadron which interacted with a nucleon is simply removed
without accounting for
a possible lowering of $z$ due to the interaction. A modification of
the string fragmentation mechanism to take this into account was
proposed in \refref{CS92}.  
Moreover, the differences at low $z$
can be ascribed to    
 target fragmentation protons, 
which  can  give a significant contribution to the multiplicity in this region.
Also the  effect of different prehadron and hadron cross sections 
$\sigma_*\neq \sigma_h$ may play a role.
This has been investigated by fixing the prehadronic state
cross-section  $\sigma_*=0.5 \, \sigma_h$. The corresponding results,
presented in the bottom right plot of \figref{fig:absresc}, show an
enhanced multiplicity ratio, but the  shape of the BG  predictions
still disagrees with the data and  the BC predictions overestimate the
experimental results. 
Therefore, we will use the BC model with  
prehadronic cross section equal to the hadronic cross section 
to predict $R_M^h$ in the region $0.2 \lesssim z
\lesssim 0.9$ also for different fragmentation products and for different
targets (see  \secref{sec:flavours}).

As we can see in \figref{fig:absresc}, the effect of the
absorption is less important at EMC in Cu 
than the absorption effect at HERMES in Kr,
which is a nucleus of similar size, due to the larger energy
transfer $\nu$ in the EMC data ($\langle \nu \rangle$=62 GeV)
compared to HERMES ($\langle \nu \rangle$$\sim$12 GeV). 
In the HERMES kinematic range both
rescaling and absorption contribute. Absorption  becomes the
dominant effect in  Kr tending to mask the rescaling effect.
In the case of the  N nucleus the theoretical result underestimates
the experimental data. This may be due to an overestimate of the
rescaling effect in light nuclei, which have a large surface to volume ratio.

It is worth to point out that 
the accuracy of the  HERMES data 
may distinguish different 
models for the space-time development of the fragmentation 
functions. The main tool in this respect is the availability of data
at large $z$,  for both light and heavy nuclei, and for
different hadrons. This combination
explores the fragmentation process in a region which has not
been investigated before.
Obviously, our model has problems for $z>0.9$ 
to explain the data. For such large $z$L,
higher-twist processes in fragmentation,
like the direct ejection of preformed pions 
in the nucleon, may be important. We think in other regions the
data do not give a specific hint on the importance of higher-twist
effects.  
Also it may be interesting  
to extend the predictions of other models of fragmentation to the full
data set.

\section{Charge and flavour dependence}
\label{sec:flavours}
\setcounter{equation}{0}

The complete particle identification of the HERMES experiment allows us
to disentangle
the information for different hadron types. Recently, the HERMES collaboration
reported measurement of $R_M^h$ for pions, kaons, protons and
antiprotons where the 
medium effects on $\pi^+$ and  $\pi^-$ are equal, while significant
differences are observed  
between $K^+$ and $K^-$ and especially between $p$ and $\bar{p}$.
The results for individual hadron species may reveal differences in
the modification of  
$q$ and $\bar{q}$ fragmentation functions as also suggested in
\refref{hightwist}, in the  
formation time of baryons and mesons \cite{HERMES,KOPNY}, or in the
size of the hadronic and prehadronic cross sections.

\begin{figure}[tbh]
\begin{center}
\begin{minipage}[t]{15.2cm}
\begin{center}
\parbox{16cm}{
\epsfig{figure=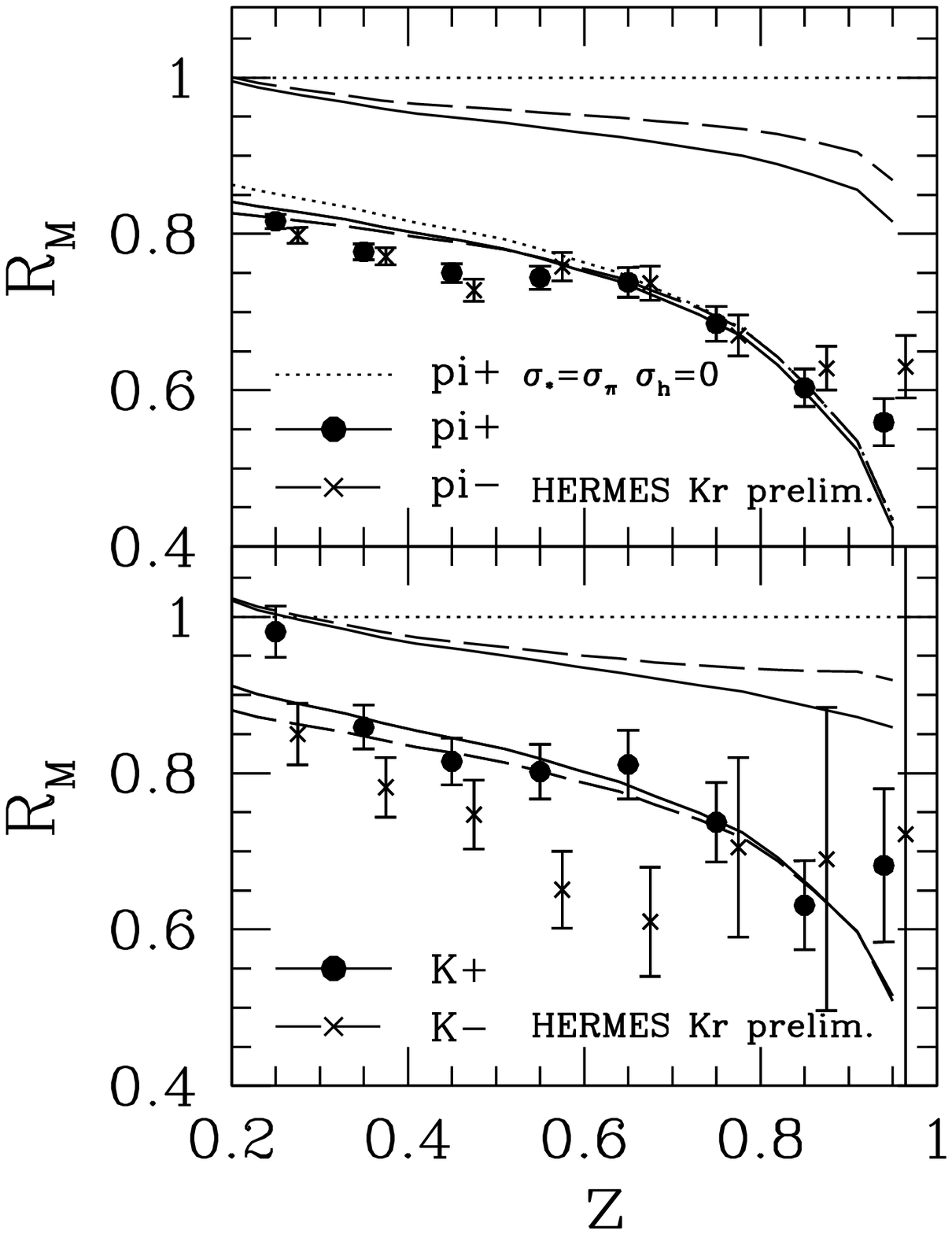,width=7.5cm
,clip=,bbllx=27pt,bblly=150pt,bburx=440pt,bbury=680pt}
\
\epsfig{figure=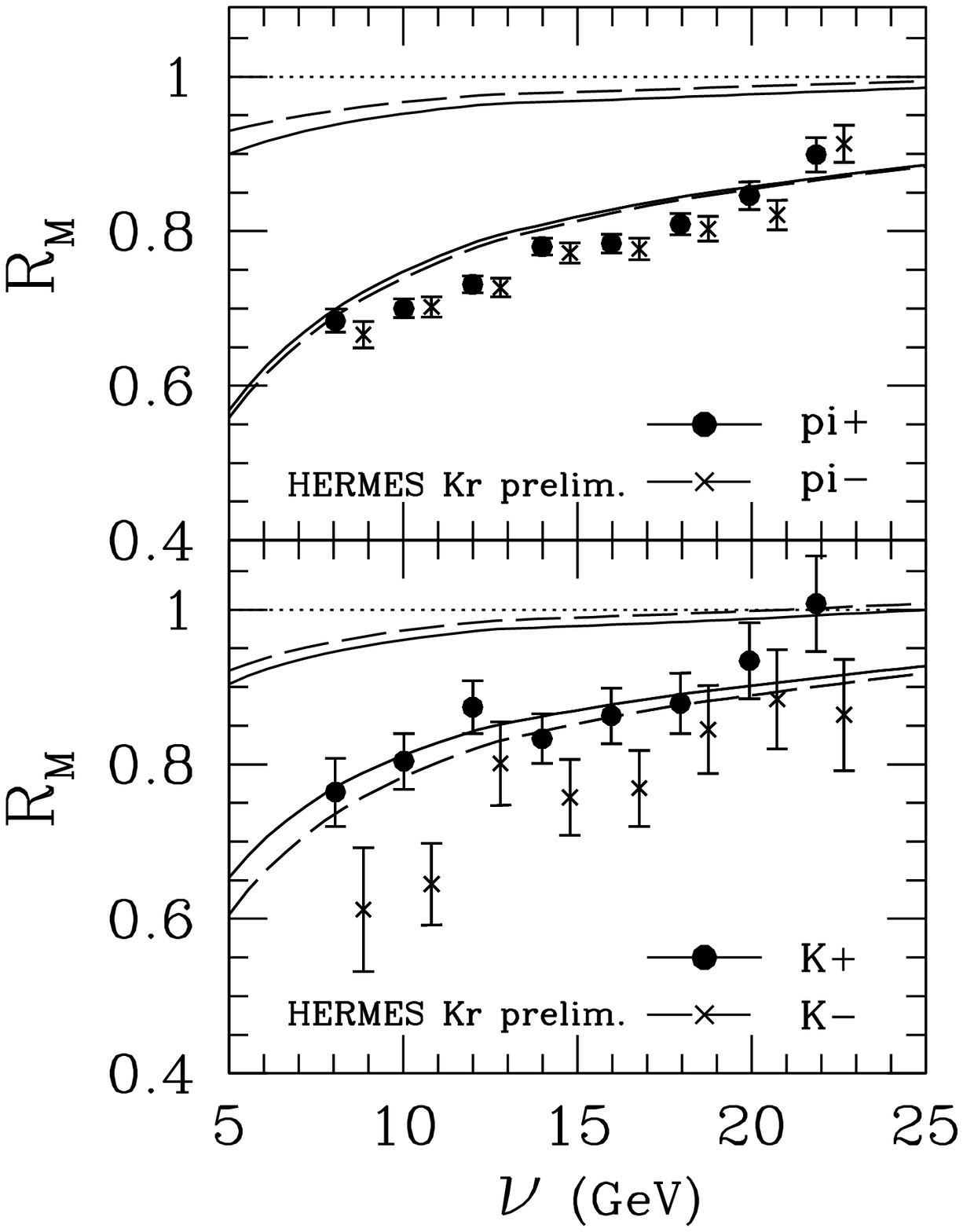,width=7.5cm
,clip=,bbllx=27pt,bblly=150pt,bburx=440pt,bbury=680pt}
}
\caption{\footnotesize
Charge- and flavour-separated theoretical multiplicity ratios 
$R_M(\nu)$ and $R_M(z)$,  compared with preliminary
HERMES data on Krypton target \cite{DATAHERMES} (the data for
negatively charged hadrons have been slightly shifted to the right to
improve the readability of the figure). 
The upper pair of curves includes
rescaling without absorption for positive and negative particles, 
and the lower pair rescaling plus
BC absorption  for positive and
negative particles. The dotted curve in the upper left plot 
show the result  
 by setting $\sigma_h$=0 and   $\sigma_*$=$\sigma_{\pi}$ in
the calculation of $R_M^{\pi}$.
The error bars represent the statistical uncertainty only.
Systematic uncertainty is about 4\% \cite{DATAHERMES}.}
\label{fig:Hermes'02Kr}
\end{center}
\end{minipage}
\end{center}
\vskip-.4cm
\end{figure}

In the following, we present the model results at the HERMES kinematics
 for charged pions and kaons, as the present 
version of this model does not apply to baryon production.
The predictions are presented in \figref{fig:Hermes'02Kr},
and compared with the HERMES data, both as a function of $\nu$ and $z$.
In each plot we show the contribution of the rescaling and the total
 contribution of rescaling and absorption for both charged states. 
One sees 
that the average charged meson spectra $\frac{1}{2}$($\pi^++\pi^-$)
and $\frac{1}{2}$($K^++K^-$) are reduced by the rescaling of the
singlet fragmentation functions, which are strongly affected by
the anomalous dimension of gluon splitting. At large $z$, the observed 
$\sim$10\% effect can be 
calculated analytically. The difference of  ($\pi^-$-$\pi^+$)
and  ($K^-$-$K^+$) spectra is governed by the non-singlet fragmentation function,
which is weighted with the excess of neutron number (N=48)
compared with the proton number (Z=36) in krypton. This explain the
slight enhancement of $\pi^-$ and $K^-$ spectra at large $z$ in the
rescaling model shown in \figref{fig:Hermes'02Kr} (upper curves in
each plot).
Subsequent absorption with a larger nuclear cross sections for
$\pi^-$ and $K^-$  (see \figref{fig:totxsec}) 
corrects this difference. 

The data
for $\pi^+$ and $\pi^-$ production indeed show no difference 
in the multiplicity ratio, and are in nice agreement with the model
predictions. We do not show calculations of $\pi^0$ mesons since they are very
similar to the curve for charged pions. 
It is worth to point out that there is a nice agreement between model
predictions and pion data also in the low $z$ region. This finding 
suggests that the observed difference in the spectra of charged
hadrons (see  \figref{fig:absresc})
should be ascribed  to  proton production from target 
fragmentation which has  not been considered theoretically.
It is also possible to disentangle the contributions of the
prehadronic and hadronic 
absorption by setting $\sigma_h$=0 and   $\sigma_*=\sigma_{\pi}$ in
the calculation of $R_M^{\pi}$. The resulting  $z$-distribution of
charged pions is shown as dotted curve in the upper left plot of
\figref{fig:Hermes'02Kr}. The
neglect of hadronic absorption leads to a very small increase of
$R_M^{\pi}$ at small $z$. This result shows that in the HERMES kinematic
region mainly the prehadron absorption contributes because the 
 hadron average formation length $\vev{l_h}$=$\vev{l_F}$+$\frac{z\nu}{k_A}$
is larger than the size of the krypton nucleus.

For $K^+$  production there is also a general agreement between data 
and model computations, while experimental results
point to a stronger than calculated absorption of $K^-$ mesons.
We found that an increase of $\sigma_*$ of at least 50\% would be necessary 
in order to reproduce the experimental data for $K^-$. The discrepancy between
theory and data may also point to  a different 
formation mechanism  for the negative kaons, as they do not contain 
any nucleon valence quarks, which dominate in the HERMES
kinematics. This may imply a shorter
formation length than predicted by \eqeqref{Lundaverlf}, 
hence a larger absorption, 
since rank 1 hadrons would not participate in the $K^-$ formation, see
\eqeqref{averlfgeneral}.

\begin{figure}[tb]
\begin{center}
\begin{minipage}[t]{15.2cm}
\begin{center}
\parbox{16cm}{
\epsfig{figure=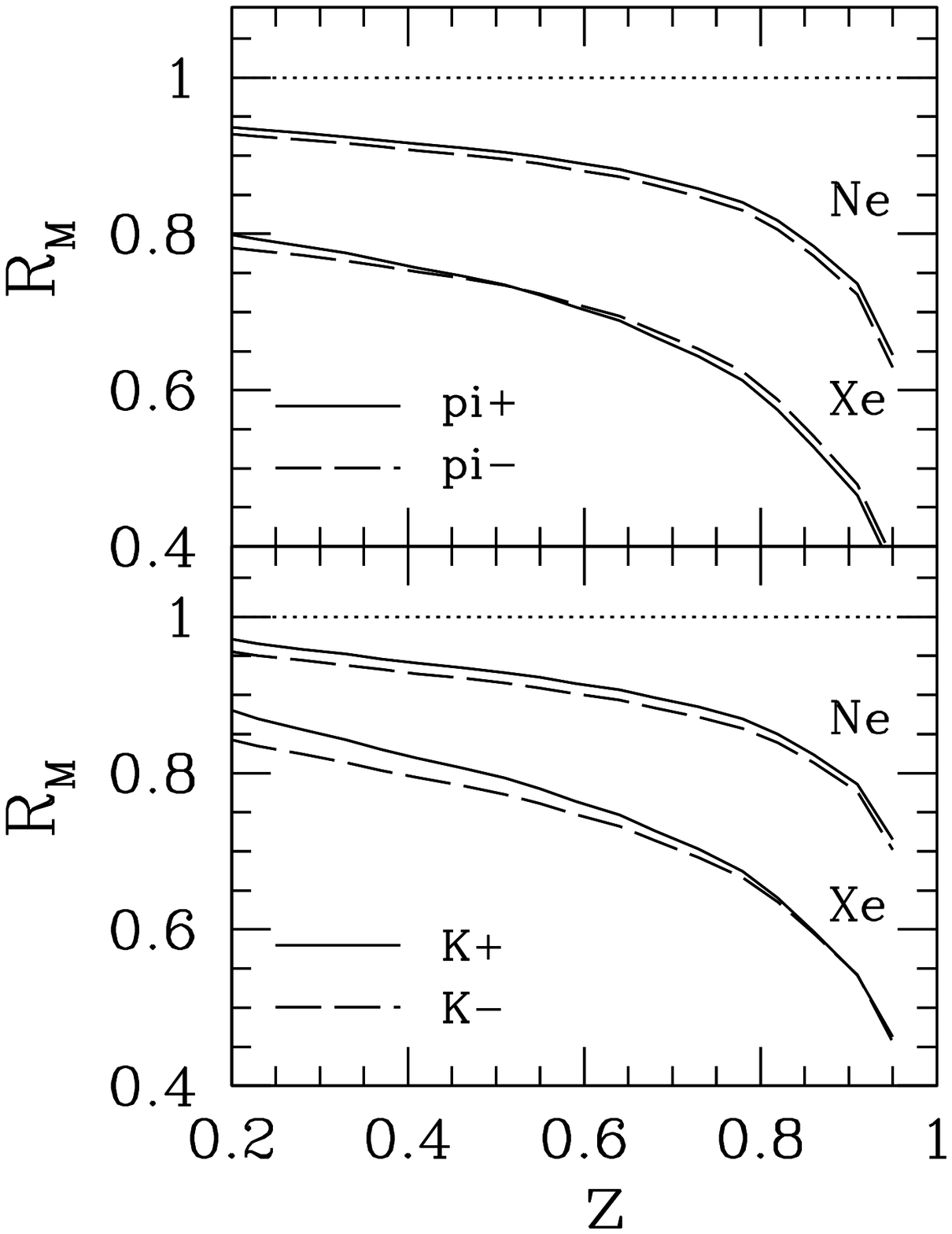,width=7.5cm
,clip=,bbllx=27pt,bblly=150pt,bburx=440pt,bbury=680pt}
\
\epsfig{figure=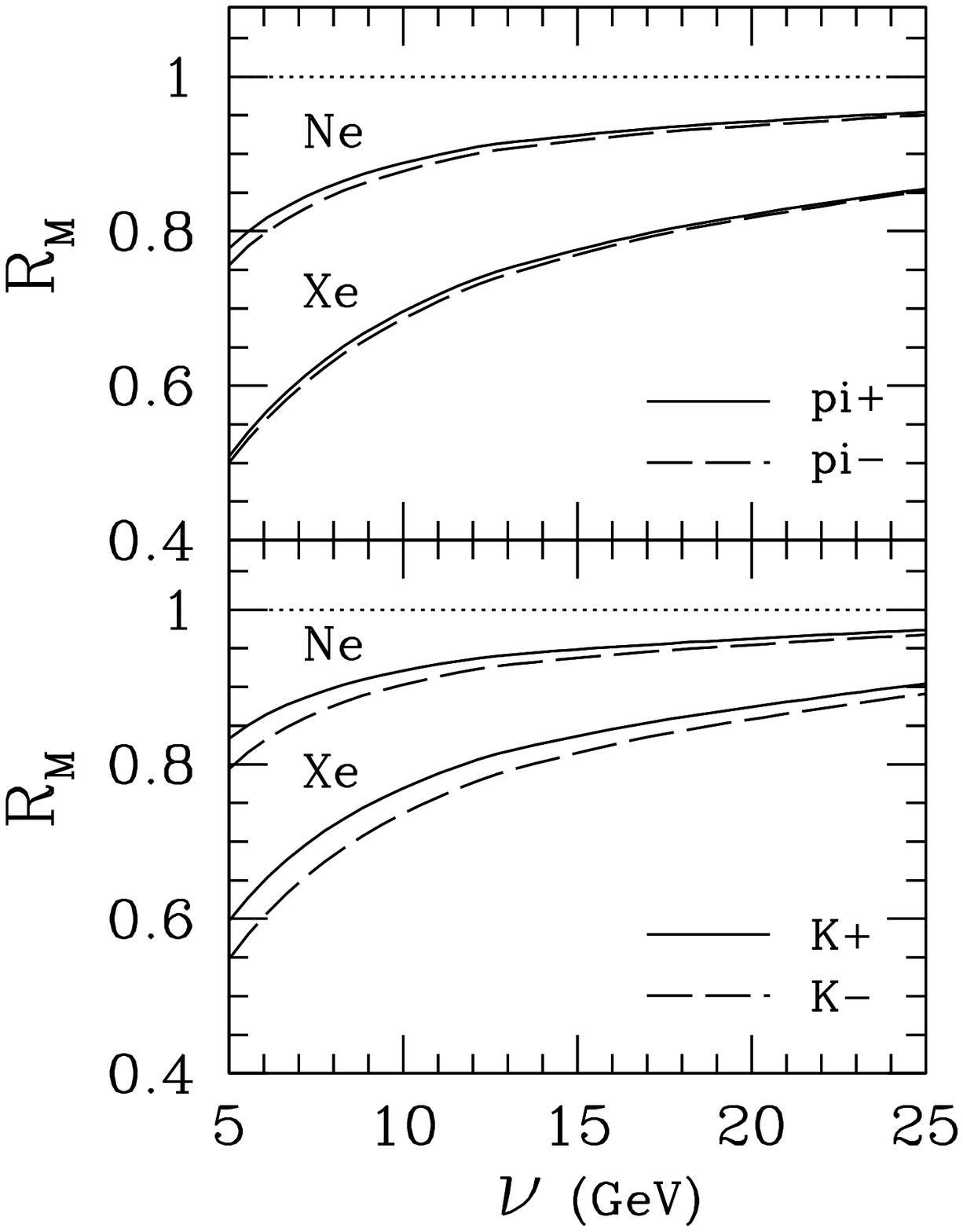,width=7.5cm
,clip=,bbllx=27pt,bblly=150pt,bburx=440pt,bbury=680pt}
}
\caption{\footnotesize
Predictions of the rescaling plus absorption  effects at HERMES 
for  charge-
and flavour-separated multiplicity ratios 
on Neon and Xenon targets. }
\label{fig:Hermes'02NeXe}
\end{center}
\end{minipage}
\end{center}
\vskip-.4cm
\end{figure}

The forthcoming HERMES data for different hadron types on light and
heavy nuclei may help to further disentangle rescaling and absorption
effects and to clarify details of the fragmentation process. 
In  \figref{fig:Hermes'02NeXe}
we show the predictions of the model for Ne and Xe nuclei
in the HERMES kinematic region for charged pions and kaons.
To this respect, the absorption
mechanism can be studied on heavy nuclei, while light nuclei are more
sensitive to the rescaling effect alone.
One characteristic feature of the rescaling corrections is 
the increase of the multiplicity ratio $R_M>1$ for $z<0.2$ which in the 
case of $K^+,K^-$ fragmentation is visible in the 
calculation with rescaling alone
as shown in \figref{fig:Hermes'02Kr}.
In addition, due to the smaller $K^+$-nucleus interaction cross section,
the   $K^+$ production on light nuclei may leave a chance to observe
the pure rescaling effect. 

\section{Discussion and Conclusions}
\label{sec:conclusions}
\setcounter{equation}{0}

We have considered two effects to describe semi-inclusive hadron
production in nuclei: gluon radiation and nuclear absorption. In the
final theoretical result the two effects add and until now the
experimental situation does not clearly identify the rescaling
effect. 
Theoretically, the gluon radiation occurs through the whole hadronization
process even outside of the nucleus and criticism may be raised whether 
the rescaling correction is inappropriate in this case.
We do not believe so.
Rescaling extends the gluon window in $Q^2$ toward the infrared
and includes thereby those gluons which are later absorbed 
by target fragments related to the nucleus. 
The fast forward gluons which go into
the current fragmentation region have a  higher time-like virtuality and 
are taken into account in the process of absorption of the prehadron.
Therefore we think that no special necessity arises to truncate  the
rescaling correction due to formation outside of the nucleus.
Of course, by assuming only a constant rescaling factor over the
whole nucleus, we  do not consider the escape of quarks in surface
nucleons directly into the vacuum. In this case a more sophisticated
treatment may be needed.

We have shown that rescaling models together with nuclear absorption 
are  able to describe both HERMES and EMC data on the nuclear
modification of hadron production in DIS 
process. While maximal deconfinement is ruled out by the
data - it assumes a too large deconfinement - partial deconfinement
is shown to have a possible effect on fragmentation functions. 
Nuclear absorption of the preformed hadron is shown to significantly
affect  the hadron production on heavy nuclei in the kinematic region of 
the HERMES experiments.
The gluon-bremsstrahlung model of \refref{KNP96}, 
that shares some similarities with the ``outside-inside'' picture of the Lund
model, is also in good agreement with data 
at $z>0.5$.
Further precise data at moderate and high $\nu$'s and for light and
heavy targets are needed to disentangle rescaling and formation length
effects and to establish rescaling as an independent mechanism on its
own.

Different scenarios for the space-time development of 
hadronization may be usefully tested against HERMES data.
In our theoretical description they enter through the
probability distributions $\cal{P_*}$, and eventually ${\cal P}_h$
if one separates between prehadron and hadron,
and through the corresponding formation lengths $\vev{l_*}=\vev{l_F}$
and $\vev{l_h}=\vev{l_F}+z\nu/\kappa_A$.
It would be interesting to measure $z$-distributions with 
low- and high-$\nu$ 
cuts in order to study how much the prehadronic
absorption differs from hadronic absorption  
as the experimental data on $K^-$ may suggest. 
This would allow to change the characteristic length
scale $L=\nu/\kappa_A$ of the hadronization process.
The high-$\nu$ data sample would be sensitive only on prehadron
absorption, because the hadron is formed almost entirely outside  the
nucleus. On the other hand, in the low-$\nu$ data sample 
the information obtained on the prehadron absorption may be used to
pin down the hadron absorption.
Furthermore, the use of different cuts on $z$ would allow
to  map out the $\nu$-dependence of the absorption
process more neatly. The necessity to have very accurate data in a
wide range of $\nu$ and with different nuclei
has also been stressed in the recommendation letter for a new European
electron facility \cite{NEEF}. 

\vskip.7cm
{\footnotesize
{\bf Acknowledgments.} 
We are grateful to N.~Bianchi, P.~di~Nezza, M.Gyulassy,
J.~H\"ufner, 
B.~Kopeliovich, A.~Metz and P.~Mulders for stimulating discussions.
This work is partially funded by the European 
Commission IHP program under contract HPRN-CT-2000-00130.
}


\newpage

\noindent {\LARGE \bf Appendices}

\begin{appendix}

\section{EMC and HERMES acceptance}
\label{app:EMCandHERMES}
\setcounter{equation}{0}

\begin{figure}[th]
\begin{center}
\epsfig{figure=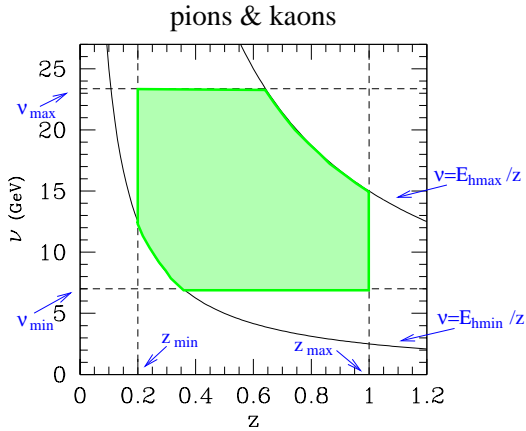,width=7.5cm
,clip=,bbllx=0pt,bblly=100pt,bburx=600pt,bbury=570pt}
\vskip-.5cm
\caption{\footnotesize
The HERMES kinematic acceptance in $z$ and $\nu$ for pions and
kaons.}
 \label{fig:kincuts}
\end{center}
\vskip-.4cm
\end{figure}

In \tabref{tab:HERMES'01cuts} we give the list of the EMC and HERMES
kinematic range and experimental cuts,  used in the numerical computations
of this paper. The HERMES kinematic acceptance in $z$ and $\nu$ for pions and
kaons is shown in \figref{fig:kincuts}. This figure explains the step
structures of the LO computation of $\vev{\nu}(z)$ and $\vev{z}(\nu)$ shown in
\figref{fig:expave}.

\begin{table}[htb]
\begin{center}
\small
\begin{tabular}{ll||l|l||l|l||} 
 & & \multicolumn{2}{|c||}{\bf EMC} & \multicolumn{2}{|c||}{\bf HERMES} \\
\hline
                 &         & \multicolumn{2}{|c||}{$h$}    
                           & $h$        
                           & $\pi,K$ \\\hline
 $E_{\rm beam}$  & GeV     & 100    & 200
                           & 27.5
                           & 27.5 \\
 $Q^2_{\rm min}$ & GeV$^2$ & 2      & 2         
                           & 1       
                           & 1 \\
 $W^2_{\rm min}$ & GeV$^2$ & 4      & 4        
                           & 4
                           & 4 \\
 $y_{\rm max}$   &         & 0.85   & 0.85    
                           & 0.85    
                           & 0.85   \\\hline
 $x_{\rm min}$   &         & 0.02   & 0.02         
                           & 0.06    
                           & 0.02    \\
 $x_{\rm max}$   &         & 1      & 1        
                           & 1       
                           & 1 \\
 $z_{\rm min}$   &         & 0.2    & 0.2    
                           & 0.2     
                           & 0.2 \\
 $z_{\rm max}$   &         & 1      & 1 
                           & 1       
                           & 1 \\
 $\nu_{\rm min}$ & GeV     & 10     & 30        
                           & 7       
                           & 7 \\
 $\nu_{\rm max}$ & GeV     & 85     & 170        
                           & 23.4  
                           & 23.4    \\
 $E_{h \rm min}$ & GeV     & 3      & 3    
                           & 1.4     
                           & 2.5     \\
 $E_{h \rm max}$ & GeV     & 85     & 170 
                           & 23.4  
                           & 15.0    \\\hline
\end{tabular}
\parbox[t]{10cm}{
\caption{\footnotesize
Kinematic cuts of the EMC and 
HERMES experiments.}
\label{tab:HERMES'01cuts}}
\end{center}
\end{table}

\section{Lund formation lengths}
\label{app:Lundcomps}
\setcounter{equation}{0}

We review  the probability distribution of the
formation lengths  of a hadron with fractional
momentum $z$ in the Lund model \cite{BG87}:
\begin{align}
    {\cal P}_h(y;z,L) = \frac{\sum_{n=1}^\infty D_n(y;z,L)}
        {\sum_{n=1}^\infty \int dy \,D_n(y;z,L) } \ , 
  \label{Phdef}
\end{align}
where $D_n(y;z,L)$ is the   longitudinal phase space 
density for the $n$-th rank hadron with fractional momentum $z$ 
to be formed at position $y$. For a prehadron with fractional
momentum $z$, the probability distribution of the formation length 
is:
\begin{align}
    {\cal P}_*(y;z,L) = \ {\cal P}_h(y+z L; z,L) \ .
  \label{P*def}
\end{align}

Consider \figref{fig:space-time}. The $n$-th pair production point $C_n$ has the light cone 
coordinates $y^\pm_n = t_n \pm y_n$.
The $n$-th rank hadron has light-cone
momenta $p^\pm_n=E_n \pm p_n$, and the fractional momenta are:
\begin{align}
    z_n = \frac{E_n}{E_0} =\frac{p_n^+}{p_0^+} \ . 
  \label{fracmom}
\end{align}
The first constituent of the
hadron $h_n$, formed in $C_n$, accelerates to the right 
under a constant string tension
$\kappa_A$. At $P_n$, 
it has a light-cone momentum $p^+ = 2\,\kappa_A\,(y_{n-1}-y_n)$.
The second constituent of $h_n$, created in $C_{n-1}$,  
accelerates to the left, and at $P_n$ it has $p^+ = 0$.  
\begin{align}
    p_n^+ = & \kappa_A \left( y^+_{n-1} - y^+_n \right) \\
  \label{pfny} 
\end{align}
The
hadron fractional momentum $z_n$ for $n\geq 1$ is determined by the
coordinates of the pair creation points (recall $\kappa_A/(2 \nu)=1/(2 L)$ 
and $y_0^+=2 L,z_0=1$): 
\begin{align}
    z_n = \frac{y^+_{n-1}-y^+_n}{2L} \ .
  \label{zfny}
\end{align}
The probability that a quark fragments into $n$ hadrons with
fractional momenta $z_1,z_2,...z_n$ is related to
the normalized string fragmentation function 
$f(z)$ which gives the probability that a string breaks into two
pieces carrying fractional momenta $z$ and $(1-z)$
\begin{align*}
     F(z_1,z_2,...,z_n) \prod_{i=1}^n dz_i = \prod_{i=1}^n f(u_i) du_i \ ,
\end{align*}
with relative fractional momenta $u_i$:
\begin{align}
    u_n = z_n \frac{1}{1-\sum_{k=1}^{n-1}z_k} \ .
  \label{defui}
\end{align}
From \eqeqref{zfny} for $y^+_i$ and using \eqeqref{defui} we
obtain: 
\begin{align*}
    y_n^+ = 2 L \left( 1 - \sum_{k=1}^n z_k \right)
        = 2 L \prod_{k=1}^n (1-u_k) \ .
\end{align*}
The longitudinal phase space densities   $D_1(y;z,L)$ of the
first rank hadron and $D_n(y;z,L)$ of the
$n\geq 2$-th rank hadron to be formed at $y$ with 
fractional momentum $z$ are:
\bea
  D_1(y;z,L) &=& \int du_1 \, f(u_1) 
        \delta(z-z_1) \delta(y-L)\\
  D_n(y;z,L) &=& \int \prod_{k=1}^n du_k \, f(u_k) 
        \delta\big(z-z_n\big) \delta\big(y-\frac12
        \left[y^+_{n-1}-y^-_n\right]\big)
        \ .
  \label{distrgen}
\eea
Since the $n$-th hadron production happens near the positive
light-cone $y^-_n\approx 0$, we can use
\begin{eqnarray*}
  \delta(z-z_n) \delta(y-\frac12 \left[y^+_{n-1}-y^-_n\right]=
  \delta(z-u_n y/L) \delta(y-L \prod_{k=1}^{n-1} (1-u_k)) 
\end{eqnarray*}
to integrate  $D_n$ over $u_n$ in \eqeqref{distrgen}. We obtain:
\begin{align}
    D_n(y;z,L) = \frac{L}{y} f(\frac{z L}{y}) \rho_n(y,L) \ , 
 \label{Dnrhon}
\end{align}
where
\begin{eqnarray}
    \rho_1(y;L) &=& \delta(y-L)\\
    \rho_n(y;L) &=& \int \prod_{k=1}^n du_k \, f(u_k) 
        \delta(y-L \prod_{k=1}^{n-1} (1-u_i))
  \label{rhon}
\end{eqnarray}
are the densities of formation points of the $n$th-rank hadron.
The total density of
hadron formation points is:
\begin{align}
    \rho_h(y;L) = \sum_{n=1}^{\infty} \rho_n(y,L) 
        = \delta(y-L) + \sum_{n=2}^{\infty} \rho_n(y,L) \ . 
  \label{rhoh}
\end{align}
Consider now the string fragmentation function of the Lund model
\cite{LundPhysRep}:
\begin{align}
    f(z) = \sum_q p_q (1+C_q) (1-z)^{C_q} \ ,
  \label{Lundfragfn}
\end{align}
where $p_q$ is a probability that a given production point involves pair production of a particular flavor $q$,
and $C_q$ are parameters controlling the typical momentum fraction carried by final hadrons containing that flavor:
\begin{align*}
    C_q = \left\{ \bay{ll}
        C              &  q=u,d,s \\
        D_q            &  q \neq u,d,s
    \eay \right. \ ,
\end{align*}
In the standard Lund model $C\approx$ 0.3
and $D_q$ is flavour dependent. Note that the subscript
$q$ includes not only quark flavours, 
but also diquarks in the fragmentation of the string
into baryons.
Since $p_q~\ll~1$ for non light flavours $q \neq u,d,s$, we obtain from 
\eqeqref{rhon} and \eqeqref{rhoh}:
\begin{align*}
    \rho_h(y,L) \approx \delta(y-L) + \frac{1+C}{y} \ .
\end{align*}
Finally, we obtain the probability distributions of the formation
length of the prehadronic state, which for a 
flavour $q=u,d,s$ has $C_q=C$ 
and for $q \neq u,d,s$ has $C_q=D_q$:
\begin{align}
    {\cal P}_*^q(y;z,L) \ = \ &
        \frac{1+C_q}{(1+C)(1+z\frac{Cq-C}{1+C_q})}
        \frac{zL}{y-zL}
        \left[ \frac{y}{(y+zL)(1-z)} \right]^{C_q} 
        \nonumber \\
    & \times \left\{ \delta[y-(1-z)L] 
          + \frac{1+C}{y-zL} \theta[(1-z)L-y]\right\} \theta[y] \ .
 \label{probdistgeneral} 
\end{align}
Note that the $\delta$-function represents the contribution from rank-1
hadrons. 
The average formation length for the prehadronic state is then
computed as $\vev{l_*}(z,L) = \int dy \, y\, {\cal P}_*(y;z,L)$:
\begin{align}
  \vev{l_*} \ = \ & \frac{1+C_q}{1+C+(C_q-C)z} 
    \nonumber \\
  & \times
       \left[ 1 + \frac{1+C}{2+C_q}\,\frac{(1-z)}{z^{2+C_q}}\,
       {}_2F_1\Big( 2+C_q,2+C_q;3+C_q;\frac{z-1}{z} \Big)
       \right]  (1-z)\, z\, L \ ,
 \label{averlfgeneral}
\end{align}
where ${}_2F_1$ is the Gauss' hypergeometric function \cite{Magnus}. 
Subtracting 1 from the square brackets in \eqeqref{averlfgeneral} would give
the average formation length for hadrons of rank $n\geq 2$.
For a valence quark, $q=u,d,s$, the above expression reduces to
\eqeqref{Lundaverlf}, which is the one used in this paper. 
On the other hand, using \eqeqref{averlfgeneral} with $C_q$=1
and $C$=0 as considered in \refref{PAVEL}, the prehadron formation length is  
\begin{align*}
     \vev{l_*} = \left[ \frac{\ln(1/z^2)-1+z^2}{1-z^2} \right] z L \ ,
\end{align*}
which is larger than in the standard Lund model as shown in
\figref{fig:averagelf}.

\section{Absorption formulae with two length scales}
\label{app:twoscales}
\setcounter{equation}{0}

When considering a different cross-section for the hadron and the
prehadron interaction with the nucleons, $\sigma_* \neq \sigma_h$,
equations \eqref{BG-W0} and \eqref{BCabs} have to be modified
to take into account the propagation of the prehadronic state.
In the Lund model considered in this paper, it 
propagates through a fixed length equal to $z\nu/\kappa_A$ before the
formation of the hadron.
 
In the Bialas-Gyulassy model \eqeqref{BG-W0} becomes
\begin{align}
    W_0(y\,',z,\nu) = 1 
        - \sigma_* \int_{y\,'}^{y\,'+z\nu/\kappa_A} \hspace*{-.2cm}dy_*\,
        \rho_A(b,y_*)
        - \sigma_h \int_{y\,'+z\nu/\kappa_A}^\infty \hspace*{-.2cm}dy_*\,
        \rho_A(b,y_*) \ .
 \label{BG-W0-2scales}
\end{align}

In the Bialas-Chmaj model, the two-stage fragmentation process
is modeled as a chain decay with two average decay lengths: 
$\vev{l_*} = \vev{l_F}$ for the ``decay'' of the quark into the
prehadronic state, and $\vev{\Delta l} = z\nu/\kappa$ for the ``decay'' of the
prehadronic state into the observed hadron ($\vev{l_F}$ is given in
\eqeqref{Lundaverlf}). \eqeqref{BCabs} becomes then
\begin{align}
    S_A(b,y) = 1 
        - \sigma_* \int_y^\infty \hspace*{-.2cm}dy\,'\,
           P_*(y\,'-y)  \rho_A(b,y\,')
        - \sigma_h \int_y^\infty \hspace*{-.2cm}dy\,'\,
           P_h(y\,'-y)  \rho_A(b,y\,')\ , 
 \label{BCabs-2scales}
\end{align}
where $P_*(y\,'-y)$ and $P_h(y\,'-y)$  are, respectively, the probability of
survival of a prehadron and a hadron at a distance $ y\,'-y$ from the virtual
photon interaction point. When $\vev{l_*}\neq\vev{\Delta l}$ one has:
\begin{align}
    P_*(y\,'-y) & = \frac{\vev{\Delta l}}{\vev{l_*}-\vev{\Delta l}}
        \left(e^{\,-(y\,'-y)/\vev{l_*}} - e^{\,-(y\,'-y)/\vev{\Delta
        l}} \right)  \ , \\
    P_h(y\,'-y) & = 1 - e^{\,-(y\,'-y)/\vev{l_*}} -  P_*(y\,'-y) \ .
\end{align}
When $\vev{l_*}=\vev{\Delta l}$ the prehadron survival probability is:
\begin{align}
    P_*(y\,'-y) & = \frac{1}{\vev{l_*}} \, (y\,'-y) \, e^{\,-(y\,'-y)/\vev{l_*}}
      \ .
\end{align}

\end{appendix}


\end{document}